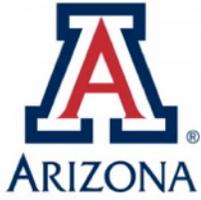
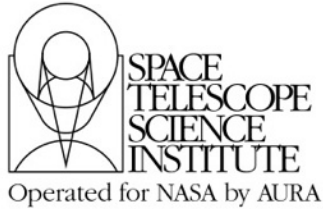

# Enabling Narrow(est) IWA Coronagraphy with STIS BAR5 and BAR10 Occulters

Glenn Schneider[1], Andras Gaspar[1], John Debes[2], , Theodore Gull[3],
Dean Hines[2], Daniel Apai[1], George Rieke[1]

[1] Steward Observatory and the Department of Astronomy, The University of Arizona, Tucson, AZ
[2] Space Telescope Science Institute, Baltimore, MD
[3] (emeritus) NASA/Goddard Space Flight Center, Greenbelt, MD

May 10, 2017

**ABSTRACT**

*The Space Telescope Imaging Spectrograph's (STIS) BAR5 coronagraphic occulter was designed to provide high-contrast, visible-light, imaging in close (≥ 0.15") angular proximity to bright point-sources. This is the smallest inner working angle (IWA) possible with HST's suite of coronagraphically augmented instruments through its mission lifetime. The STIS BAR5 image plane occulter, however, was damaged (bent and deformed) pre-launch and had not been enabled for GO science use following the installation of the instrument in 1997, during HST servicing mission SM2. With the success of the HST GO 12923 program, discussed herein, we explored and verified the functionality and utility of the BAR5 occulter. Thus, despite its physical damage, with updates to the knowledge of the aperture mask metrology and target pointing requirements, a robust determination of achievable raw and PSF-subtracted stellocentric image contrasts and fidelity was conducted. We also investigated, and herein report on, the use of the BAR10 rounded corners as narrow-angle occulters and compare IWA vs. contrast performance for the BAR5, BAR10, and Wedge occulters. With that, we provide recommendations for the most efficacious BAR5 and BAR10 use on-orbit in support of GO science. With color-matched PSF-template subtracted coronagraphy, inclusive of a small (±1/4 pixel) 3-point cross-bar dithering strategy we recommend, we find BAR5 can deliver effective ~ 0.2" IWA image contrast of ~ $4 \times 10^{-5}$ $pixel^{-1}$ to ~ $1 \times 10^{-8}$ $pixel^{-1}$ at 2". With the pointing updates (to the PDB SIAF.dat file and/or implemented through APT) that we identified, and with observing strategies we explored, we recommend the use of STIS BAR5 coronagraphy as a fully "supported" capability for unique GO science.*

# 1. Introduction

Throughout the course of its more than 27-year long mission, the *Hubble* Space Telescope (*HST*) has hosted five instruments with coronagraphic optics for high contrast imaging. Spherical aberration in the telescope primary mirror, unrecognized pre-launch, however, rendered unusable the first two coronagraphs flown, in the first-generation instrument suite with the Faint Object Camera (FOC; f/288 channel) and the Wide-Field/Planetary Camera-1 (WF/PC; PC 8 channel). While *Hubble*'s aberrated vision was corrected individually for each instrument in Servicing Mission 1, this was not the case in the FOC f/288 channel that implemented its coronagraph. The WFPC-2 replacement instrument eliminated its predecessor's coronagraphic "Baum spot", thereby delaying the onset of *HST* coronagraphic science until the installation of the second-generation instruments in 1997 with the Near-Infrared Camera and Multi-Object Spectrometer (NICMOS) and the Space Telescope Imaging Spectrograph (STIS). NICMOS provided near-IR coronagraphic imaging from 1.1 to 2.4 μm with a symmetrical inner working angle (IWA) of r = 0.3" until its shutdown in 2009. STIS included a deployable coronagraphic focal plane mask with four hard-edge occulters (see Fig. 1). STIS provides broadband visible-light coronagraphy with:

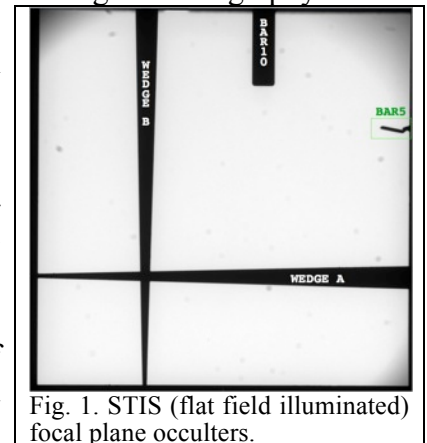

(a) two long, orthogonally oriented wedge-shaped occulters (WEDGE-A and WEDGE-B), each tapering to a minimum half-width of ~ 0.3".
(b) a 3"-wide rectangular bar-shaped occulter (BAR10) extending 10" from one edge of its 52" x 52" FOV.
(c) a, *planned*, narrow IWA bar (or "finger"-shaped) occulter (BAR5), 5" in length (orthogonal to, but non-intersecting with BAR10), with a half-width of only 0.15".

The STIS BAR5 occulter was going to provide, at visible-wavelengths, an IWA (in one dimension) half the size of NICMOS's coronagraph. Unfortunately, BAR5 was damaged pre-launch (bent and deformed) without possibility of repair or replacement.

Fig. 1. STIS (flat field illuminated) focal plane occulters.

Thus, BAR5 was not commissioned with the other three STIS coronagraphic occulters on-orbit, and was consequentially not used for subsequent science operations. The Advanced Camera for Surveys (ACS) was installed in 2004 with a high-performance visible-light coronagraph (well apodizing the *HST* diffraction spikes, unlike STIS), but with two larger angle (r = 0.9" and r = 1.8") occulters only. The unavailability of the STIS BAR5 occulter to GO programs, however, precluded the conduction of coronagraphic science investigations demanding the smallest IWAs.

In GO 12923, despite its physical damage, we explored and confirmed the utility of STIS BAR5 coronagraphy leading to its much-delayed enablement and commissioning for GO science as a supported capability. Herein we discuss the GO 12923 program and its findings leading to the restoration and contrast calibration of STIS BAR5, and BAR10-corner (§ 14), coronagraphy.

# 2. GO 12923 - Observation Plan

The *HST*/GO 12923 program was designed, with an economy of orbits (6), to explore and validate the utility of the STIS BAR 10 rounded-corner and BAR5 ("bent finger") occulters for narrow-angle coronagraphy, despite BAR5 having sustained pre-launch damage and physical deformation. Primary concerns (which were put to rest) included the possibilities of afocality of the "finger" after bending leading to contrast loss, and/or introduction of edge-defects that could induce instrumental scattering or other optical artifacts in the unocculted field beyond the finger edges. Additional concerns for robust narrow-angle coronagraphy resulting from target acquisition (placement) imprecision with a (pre-execution) highly-uncertain occulter-edge metrology were also evaluated (and retired). This included in the observing (and analysis) plan a necessary high-precision re-determination of the occulter aperture effective location in the focal plane. This necessitated multiple fine repointings (image scans) with sufficient, but only relatively very shallow depth (low SNR), imaging as would be otherwise be implemented differently with other (scientific) priority. The *HST*/GO 12923 program was conducted in three parts:

1) Pre-execution redetermination and estimation of the locations of the deformed BAR5 and supporting BAR10 aperture locations and pointing fiducials that may have changed compared to "best available" data from pre-launch design due to zero-gravity release, two decades of desorption in the OTA and instrument optical bench, and astronauts stomping around the *HST* aft shroud during various servicing missions including a major on-orbit repair to STIS itself.

2) A coarse coronagraphic step-and-dwell image scan test of the (step 1) best-assumed aperture metrology to update the BAR10 (reference aperture) to BAR5 inter-aperture metrology, develop fine corrections for then "routine" use of BAR5, and obtain a first assessment of coronagraphic viability for the use of BAR5 using the coarse image scan data.

3) Pointing corrections derived from step 2 were applied to obtain follow-on observations of a second test target, using finer imaging scans to verify the (step 2) updated BAR5 pointing and coronagraphic performance results. Most importantly, this includes the quantifying both raw and PSF-template subtracted contrast curves for BAR5 coronagraphy, and use of the BAR10 corners.

In steps 2 and 3, rather than image only point-source position and contrast calibrators, in GO 12923 we imaged two targets with well-studied edge-on circumstellar disks of well-established morphology and photometry in spatial regions commonly accessible to BAR5 and BAR10 corners beyond their unique IWA domains. This enabled an ability to test and validate with high confidence the BAR5 and BAR10 image fidelity and photometric efficacy with *a priori* knowledge in spatial regions of representative high-priority science targets previously observed.

## 3. Targets

Potential performance-degrading effects for coronagraphy can have wavelength dependencies in the unfiltered STIS 50CCD broadband response. Hence, GO 12923 was carried out with one "red" and one "blue" test target: AU Mic and β Pic, respectively (see Table 1). Both of these stars host bright, edge-on, debris disks with nearly equal visible-light scattering fractions, *a priori* characterized with STIS coronagraphy using its Wedge A and/or B occulters at their narrowest working angles (2x larger than BAR5). This allowed direct comparison of performance in regions of spatial overlap and, for BAR5, extending interior to the Wedge A/B IWA. Each test target was contemporaneously observed with a Δ|B-V| ≤ 0.04 color-matched PSF template star (see Table 1), for BAR5 and BAR10 PSF template subtraction.

Table 1 - Target and PSF Stars

|  | Name | $V_{mag}$ | B-V | V-R | Spec | $f_{disk}/f_*$[†] | Coordinates (2000.0) |
|---|---|---|---|---|---|---|---|
| Red Target | AU Mic | 8.627 | +1.423 | +0.959 | M1V | $2.5 \times 10^{-3}$ | 20h 45m 09.53s -31° 20' 27.2" |
| Red PSF | HD 191849 | 7.966 | +1.453 | +0.913 | M0V | -- | 20h 13m 53.39s -45° 09' 50.5" |
| Blue Target | β Pic | 3.86 | +0.17 | +0.12 | A6V | $2 \times 10^{-3}$ | 05h 47m 17.09s -51° 03' 59.4" |
| Blue PSF | δ Dor | 4.36 | +0.21 | +0.26 | A7V | -- | 05h 44m 46.38s -65° 44' 07.9" |

[†]$f_{disk}/f_*$ = fraction of 0.6 μm disk-scattered starlight beyond prior IWA limits (Sch14; Lagrange et al. 2014 PP-IV 639)

## 4. Orbits and Orientations

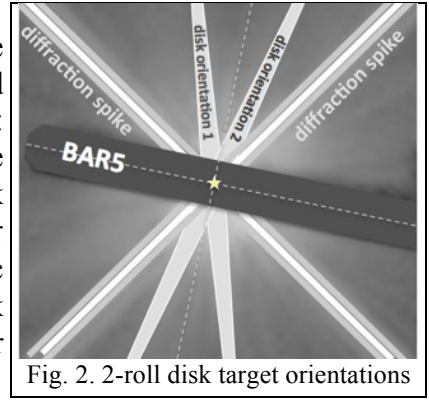

Each target and its color-matched PSF template star were observed in three contiguous orbits, with the PSF star interleaved between two target orbits each at different spacecraft roll angles: Target (roll 1)--PSF--Target (roll 2). For BAR5, the absolute orientation angles for the target orbits placed the edge-on disk straddling a line roughly perpendicular to the long axis of the bar ± 15° from roll 1 (ORIENTAT for AU Mic 295.06°, for β Pic 208.06°) to roll 2; see Fig. 2. This also cleanly placed the disk between the "upper" and "lower" sets of diffraction spikes. For BAR10 the disk celestial orientations were the same as for BAR5.

Fig. 2. 2-roll disk target orientations

## 5. BAR10-Edge and Cross Mid-BAR5 Step-and-Dwell Imaging Scans

A BAR5 aperture fiducial was not established or known to APT or other elements of the ground system when the observation plan[1] was designed (but later resulted from this test). Hence all target pointings in the observing plan were specified relative to (offset by POS TARGs from) the BAR10 aperture fiducial. By design, the latter was intended to be on the long axis of the BAR10 occulter, interior to the mask -- though with detailed examination of on-orbit flat-field images we found that not to be the case with high accuracy. Thus, for this program to ascertain the relative metrology of the BAR10 and BAR5 occulters with highest precision, we designed imaging scans to observationally determine any unintended offsets w.r.t. BAR10 rounded corners, as well as across the BAR5 occulter. For details of planning and process, see Appendix A. The AU Mic visits (with PSF star, Visit ids 04 - 06) executed first on 31 July 2013. Image-structure and pointing analysis revealed small, correctable, offsets in the commanded vs. executed positions. Details of the metrical analysis are presented in Appendix B. The residuals from these initial results were used to update/adjust the second epoch (23 Sep 2013) scans for β Pic and its PSF star δ Dor (Visit ids 01 - 03) and lead, with post-execution confirmation, to recommend Project Data Base (PDB) and derived updates for relevant STIS apertures parameters.

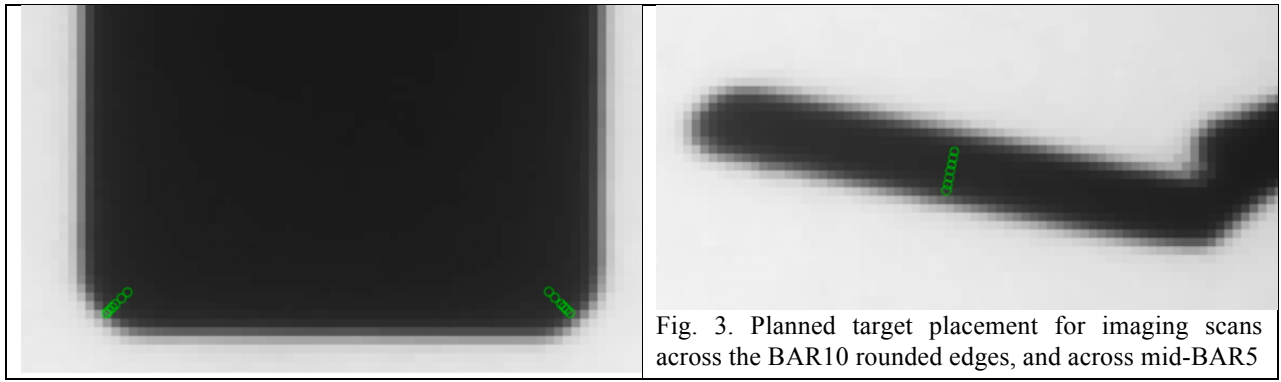

Fig. 3. Planned target placement for imaging scans across the BAR10 rounded edges, and across mid-BAR5

The planned imaging scan positions for the first-executed, AU Mic observations, as illustrated in Fig. 3, are tabulated in Science Instrument Aperture Frame (SIAF) and POSition TARGget coordinates (all relative to the SIAF and APT supported BAR10 aperture fiducial); see Table A-1.

---

[1] available at: http://www.stsci.edu/cgi-bin/get-proposal-info?id=12923

## 6. Coronagraphic Image Acquisition

Each visit began with a two-FGS guide star acquisition, followed by a standard mode-2 target acquisition exposure to place the target at the first pointing position in the visibility period. Short, individual, coronagraphic exposures were designed to approach, but not exceed, safely conservative 50% - 80% full well depth at the edge of the coronagraphic aperture, 0.15" from the core of an occulted star. Exposure times for the first epoch observations were estimated from prior coronagraphic imaging so as not to saturate. Exposure times were adjusted with both the STIS imaging ETC and TinyTim template PSFs and allowing for as much as 1/2-pixel of target acquisition decentering: for AU Mic 2.3s, for HD 191849 0.2s, for β Pic and δ Dor 0.1s (the minimum possible exposure time). At each step and dwell scan position, multiple short exposures were taken to build SNR and to mitigate effects of cosmic ray hits: 5 repeats for all targets except HD 191849 with 6 repeats. These exposure times were sufficient to enable both astrometric pointing analysis (and corrections) and contrast performance measurements while also revealing the host-star disks (though at low SNR) for comparison to existing observations obtained with other (larger) STIS coronagraphic occulting masks.

## 7. Image Calibration/Reduction

We used the *calstis* pipeline S/W with flat, bias, dark, and other calibration references files provided by STScI's calibration database system, to instrumentally calibrate the raw STIS images then transformed into count-rate images with CCD-GAIN=4 used in all observations. STIS "herringbone noise" was removed using *autofillet* (Janson et al. 2003, Proc. *HST* Cal. Workshop, 193). The multiple individual instrumentally calibrated images at each scan position then were manually inspected for any anomalies (none found) and median combined into a single count-rate image at each scan position.

## 8. Astrometric Determination of the Stellar (SIAF) Position

The SIAF location of the occulted (or partially occulted) star in each image was determined using the "X marks the spot" method that has previously been validated and used extensively in GO program 12228 (see Schneider et al. 2014, AJ 148 59; henceforth Sch14). Simply, the photometric mid-line intensity peak along the orthogonal linear OTA diffraction spikes were found, fit, and the intersection computed. Over the few arcseconds of the region considered the (very small) effects of differential geometrical distortion are negligible. For BAR5 this is very straight forward, as the diffraction spikes are symmetrically seen uninterrupted in the four diagonal directions originating at the target location. For the BAR10 corners (lower left illustrated in Fig. 4), one of the two spikes is nearly fully occulted on one side of the star by the mask itself - though it reappears in the sub-array used for data readout on the side opposite the star and is used to better constrain the fit.

Note that the eye is highly biased in assessing "where is the star?" in images (such as in Fig. 4) where part of the PSF core, with FWHM appx 1.4 pixels, is partially obscured by the BAR10 edge. The eye tends to place the star at the bright but asymmetric and partially occulted photocenter of the PSF -- which is NOT where the star is. The diffraction spikes (and origin at the true stellocenter) are unaffected by the introduction of the first focal plane BAR10 mask. For example, Fig. 4 is a BAR10 image (left; OBZE06030) that was planned to have the stellocenter 0.2" interior to the BAR10 mask along a 45° diagonal ([+2.76, -2.76] pixels in SIAF X/Y) tangent to the mask lower-left rounded corner (green dot, right panel). The "X marks the spot" solution reveals the star was not placed there, but was offset from that planned pointing by +0.068 -1.052 pixels (red dot). Details relating to that offset, and recommendations for remediation, are discussed below.

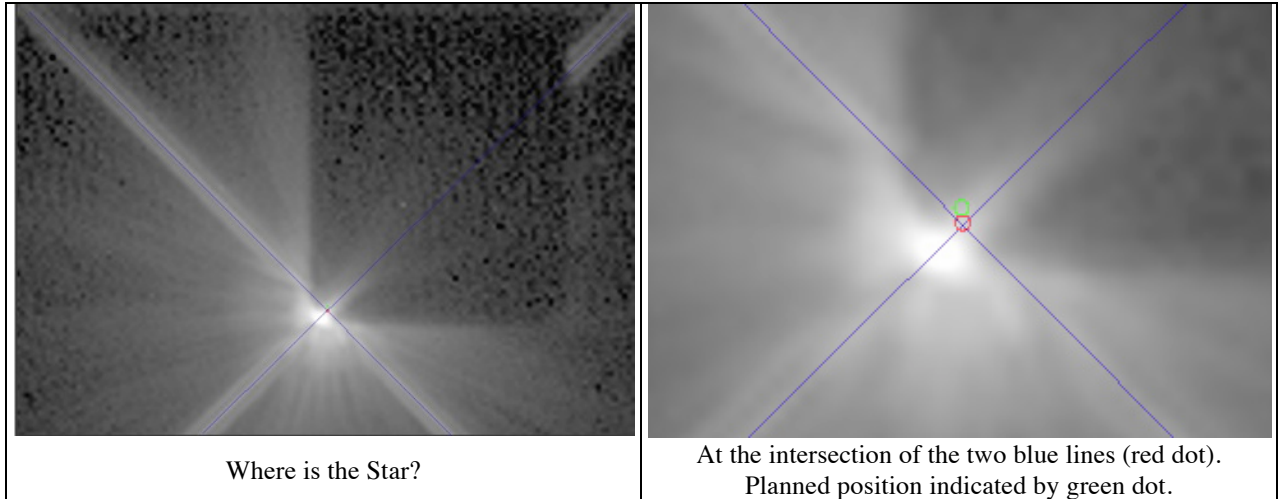

Where is the Star?

At the intersection of the two blue lines (red dot). Planned position indicated by green dot.

Fig 4. Stellar location by diffraction spike fitting "X-Marks-the-Spot" method.

Also note, the occulter edge itself is not geometrically sharp (in part because the mask plane is not perfectly conjugated to the reimaged focal plane, i.e., it is very slightly afocal when re-imaged on the detector, but also due to diffraction and scattering effects. In planning the point-position scans in this program we ascertained and used the edge-metrology from back-lit bright-sky and flat-field images at the half-power point of the brightness profile. To optimize for coronagraphic starlight suppression of the PSF core at a fixed baseline position, this 50% criterion may have been somewhat under-conservative, but was robust with the as-executed scan positions.

## 10. First Imaging Results – AU Mic; Visits 04, 05 (PSF), and 06

*10.1 Target Positioning*

Figs. B-1 through B-3, for BAR5, and the BAR10 lower-left (LL) and lower-right (LR) corners, show visit-level reduced images from each scan point presented in increasing +Y position order (incrementally smaller Y-POSTARGs from BAR10 between positions). "Visit-level" images are median combinations of all replicated images from each scan point individually calibrated with *calstis*. Figs. B-1 through B-3, are all shown with a $\log_{10}$ stretch from [+1] to [+4] dex counts $s^{-1}$ pixel$^{-1}$.

In Fig. B-1, as can be seen visually, the best centering on the mid-line of the BAR5 occurred between planned bar-perpendicular offsets of +0.04" and +0.08" in all three target visits: 04 = AU Mic (roll 1), 05 = HD 191849 PSF, 06 = AU Mic (roll2). The small red circles indicate where the star was actually placed as determined from "X-marks the spot" centroiding. Table B-1 quantitatively gives the corresponding as-planned vs. as-executed target positions and offsets derived from these images. Note that the intra-scan repointings are well correlated between all three visits. All are also consistent within an expected dispersion (of about 1/4 pixel = 0.0125") in target-to-target non-repeatability due to Mode Selector Mechanism (MSM) non-repeatability in mask repositioning with deployment.

*N. B.:* The VERY large tabulated differences in SIAF-Y is an artifact due to the fact that only small sub-array strips (100 rows wide, with offset Y-index about the regions of interest) were read out to reduce on-orbit dead time: centered at read-out row 720 for BAR5, and 840 for the BAR10 corners (Figures and Tables B-2 and B-3). The subarray start index in Y is 670 and for the BAR5, and 790 for the BAR10, images. This sub-array start index is later subtracted to get the actual difference w.r.t. the SIAF locations on the full (but not fully read-out) detector.

The intra-visit differences from planned to achieved positions were highly repeatable with very small dispersion within expectations of rms jitter for two-star fine lock (~ 4 mas = 0.08 pixels). The inter-visit mean differences have a larger dispersion, posited as due to the repeatability limits of initial target placement with independent target acquisitions (at different rolls and targets), separate from MSM deployment imprecision.

• **For BAR5** (Appendix B, Fig. & Table B-1): At each of the scan points, the target "missed" the planned position an average of (0.369, 1.366) ± (0.0577, 0.261) [1-$\sigma$] pixels. For an imaging scale of 0.05077 arcseconds pixel$^{-1}$, the as-planned BAR5 POS TARGS should have been altered by (+0.01873, +0.06935) arc seconds.

• **For BAR10 LL** (Appendix B, Fig. & Table B-2): In a similar method of analysis at each of the scan points, the target "missed" the planned position an average of (0.0388, 0.928) ± (0.069, 0.191) pixels. The as-planned BAR10 POS TARGS (for LL corner) should have been altered by (+0.00197, +0.04711) arc seconds.

• **For the BAR10 LR** (Appendix B, Fig. & Table B-3): At each of the scan points, the target "missed" the as-planned position by a very similar average of (0.0716, 0.9624) ± (0.087, 0.267) pixels. The as planned BAR10 POS TARGS (for LR corner) should have been altered by (+0.00364, +0.04886) arc seconds.

These pointing updates implicitly correct for geometrical distortions between the BAR10 aperture fiducial and the determined "best" BAR5, and BAR 10 LL & LR locations, and were applied to the β Pic and δ Dor PSF visits 01 – 03 before execution on 23 Sept 2013.

Additionally, for both the BAR10 LL and LR corners, the stellar PSF core disappears between scan step *planned* positions of +0.20" and +0.25" from the *as defined* rounded edge of the corners. We therefore adopt a "best" position mid-way between these scan points (+0.225" w.r.t round edge) as the corrected mid-point of the BAR10 scans.

*10.2 BAR5 IWA Performance (AU Mic, Visits 04-06)*

AU Mic, as imaged in both Visits 04 and 06, was very close to BAR5-centered at the step-and-dwell image scan last position (3rd exposure set in time in both visits, planned for a +0.08" offset). The PSF-template observations from Visit 05 do not precisely replicate the AU Mic visit scan offsets. However, as executed, they nearly equally flanked the near BAR5-centred AU Mic positions with the PSF pointings at positions 4 and 5 (+0.04" and +0.08" planned offsets). For these reasons, only these occulter position-matched data (AU Mic Visit 04/06 scan position 5, and PSF Visit 05 scan positions 4 and 5) were used to evaluate the BAR5 coronagraphic performance. The two BAR5 best-centered AU Mic images (one each from Visits 04 and 06) and the two most-closely corresponding PSF images from Visit 05, flanking the AU Mic positions, are shown in Fig 5.

In all cases, the unoccluded PSF halo at the closest proximity to the BAR5 edges is well exposed, but not saturated. So (in principle) the exo-BAR5 data are valid for PSF-subtraction in all regions (if not photon-limited), but are degraded in directions of the *HST* diffraction spikes.

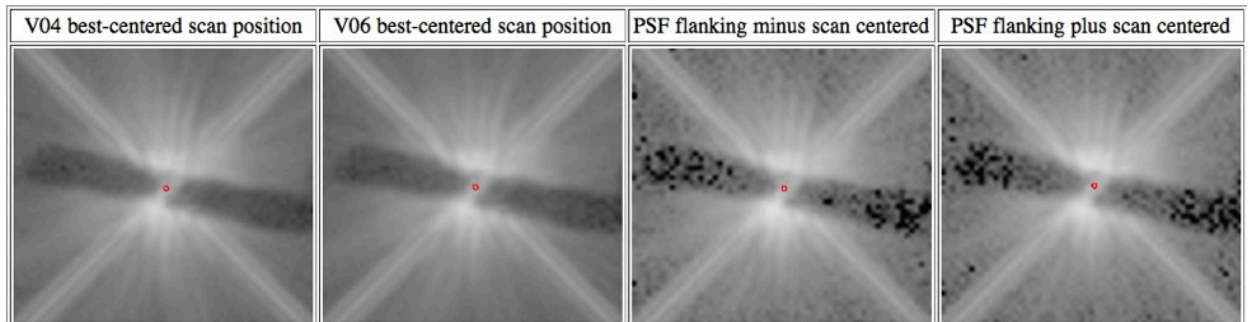

Fig. 5. AU Mic (left 2 panels, both position 5) and closest position-matched HD 191849 PSF images (right 2 panels, positions 4 and 5, respectively). All images $\log_{10}$ display dynamic range from 0.1 to 10,000 counts s$^{-1}$ pixel$^{-1}$.

Four individual PSF-template subtracted images of the AU Mic disk as shown in Fig. 6 were created from the count-rate images from each of the scan positions shown in Fig. 5:

AU Mic Visit 04 Position 5 − PSF Visit 05 Position 4
AU Mic Visit 04 Position 5 − PSF Visit 05 Position 5
AU Mic Visit 06 Position 5 − PSF Visit 05 Position 4
AU Mic Visit 06 Position 5 − PSF Visit 05 Position 5

An empirically determined PSF flux-scaling (intensity renormalization) factor of x0.58 relative to AU Mic was used. For each subtraction, the PSF template image was aligned to the location of the star in its corresponding AU Mic image initially by "X marks the spot" centroids, then with very fine empirically determined offsets minimizing the residuals in subtraction in the OTA diffraction spikes (see Sch14). The final set of the four individual PSF-subtracted images are shown in Fig. 6 (left and center panels). As previously noted for the purpose of this commissioning program, the exposure depth is sufficient to verify coronagraphic performance, but is very shallow (photon limited) in revealing the AU Mic disk. With a different (science) priority, deeper integrations (e. g., following GO 12228; Sch14) would be warranted.

As can be seen in the Fig. 6 images, the further suppression of much of the remaining starlight after BAR5 coronagraphy, by PSF-template subtraction, is both effective and highly repeatable. This is evidenced, after target/template star alignment and scaled-subtraction, by the very deep near-nulling of the OTA diffraction spikes.

Aligning the imperfectly BAR5-centered template star to the position of the imperfectly BAR5-centered target star, however, causes a mis-registration of the BAR5 edges. Indeed, the stellocenter and BAR5 edges cannot be simultaneously aligned unless both the target and template star were exactly pointed to the same place (ideally BAR5-centered). By taking out the unintended target/template small pointing offsets to null underlying light in the stellar PSF halo, the star-illuminated BAR5 edges become mis-aligned and in difference images result in opposite edge ± brightness patterns at the opposing BAR5 edges. This is easily seen in the Fig. 6 images where the disk appears at the two different celestial orientations used in Visits 4 and 6.

An inversion in parity in the ± brightness pattern at the BAR5 edges (on opposite sides of the star) is seen comparing the PSF subtracted images using the PSF templates from offset positions 4 & 5. This indicates that a better nulling of the ± pattern would have been achieved with a template image taken between these two positions. The brightness of these artifacts at the + and - edges is (to first order) "close to" equal in amplitude (but opposite in sign) in the position 4 and position 5 PSF subtracted images (though slightly different for the respective Visit 04 and 06 images). This implies that an intermediately placed PSF template, acquired approximately mid-

Preprint: Under Review                          Instrument Science Report STIS 2017-## Page   8

way between scan positions 4 and 5 should have been able to simultaneously null the PSF halo, and light scattered by the BAR5 edges. None exists for these observations. However, by digitally masking the + and − artifacts and then combining the images, we are able to produce a rudimentary science quality image (Fig. 7). The achieved IWA of this reduced image at 0.20" closely approaches the hard-edge limit of 0.15". The image scans identified the ideal POS TARG offsets from the BAR10 aperture fiducial required, and with ± 1/4-pixel cross-bar dithers to mitigate pointing uncertainties should reliably allow high quality and small IWA data acquisition (see § 15, item 5).

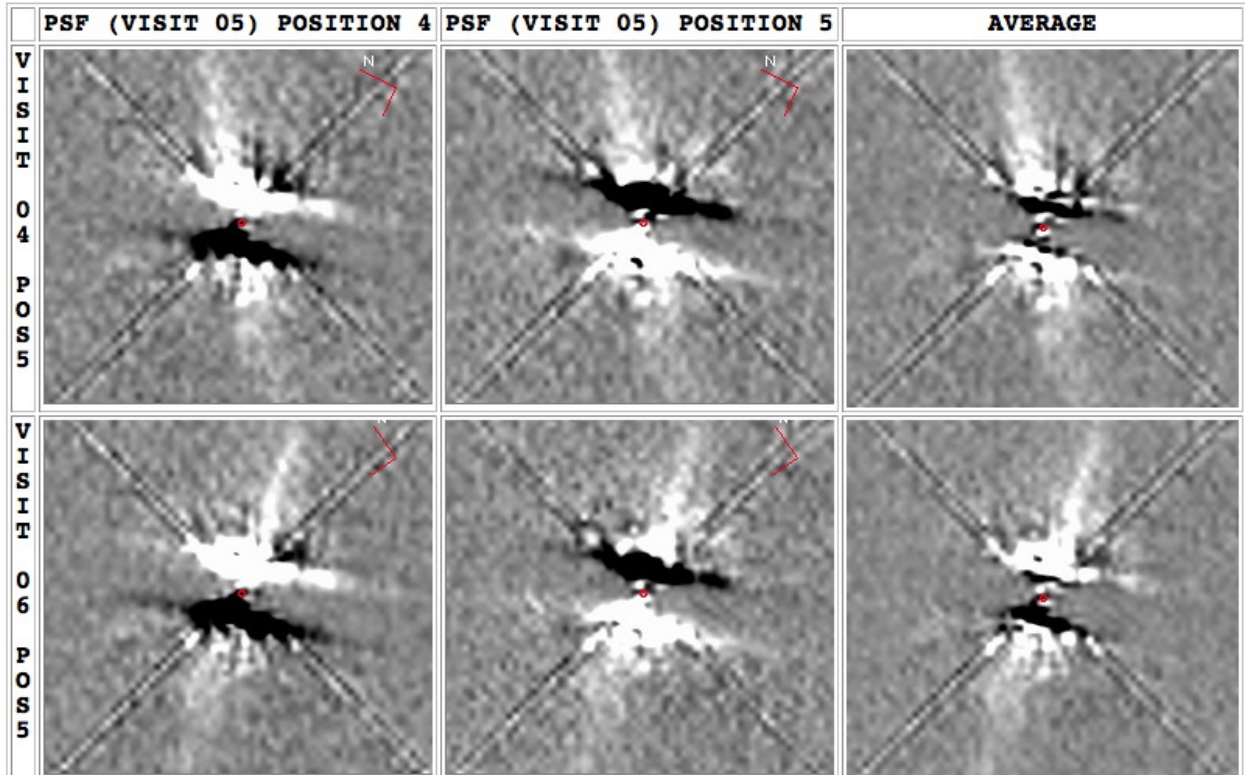

Fig 6. BAR5 PSF-template subtracted images of the AU Mic inner disk. Top: 1st roll (Visit 4) using PSF templates from two adjacent scan positions. Bottom: 2nd roll (Visit 6) using same PSF templates. Linear display with symmetric ± 20 counts s$^{-1}$ pixel$^{-1}$ stretch to show PSF subtraction residuals as well as light from the disk. (Compare starlight suppression w.r.t. Fig. 5 $\log_{10}$ display maximum at 10,000 counts s$^{-1}$ pixel$^{-1}$).

Note: These AU Mic data preceded the Visit 01 - 03 β Pic data by ~ 2 months. For those later β Pic observations, along with the initial pointing correction offsets previously discussed applied, finer scan spacings were implemented and compared to the coarser "out of the box" AU Mic observations. The pointing corrections applied were based upon the above nearly nulled positions.

To first order in this data set, the phase reversal of this ± BAR5-edge artifact brightness pattern from Visit 04 to 05 suggested that, as a proxy to having an "intermediate" PSF template, applying linear combination of the two available PSFs could reduce the magnitude of this pollutant. This was done, for simplicity here just assuming equal weighting for the two images, with averaged results shown in the third column in the above mosaic of images.

The two averaged images of the AU Mic disk at different celestial orientation angles (Fig. 6. right panels), were then "roll combined". I.e., co-aligned at the location of the star, rotated to celestial north "up", and then averaged with digital masking (rejection) of regions unsampled or degraded by the presence of the BAR5 occulter and the *HST* diffraction spikes. This two-roll combined image is shown in Fig. 7, clearly detecting and resolving the AU Mic disk along its edge-on major axis to a smallest stellocentric angle of r = 0.20", closely approaching (less than an innermost resel[2] from) the 0.15" half-width of physical edge of the BAR5 aperture.

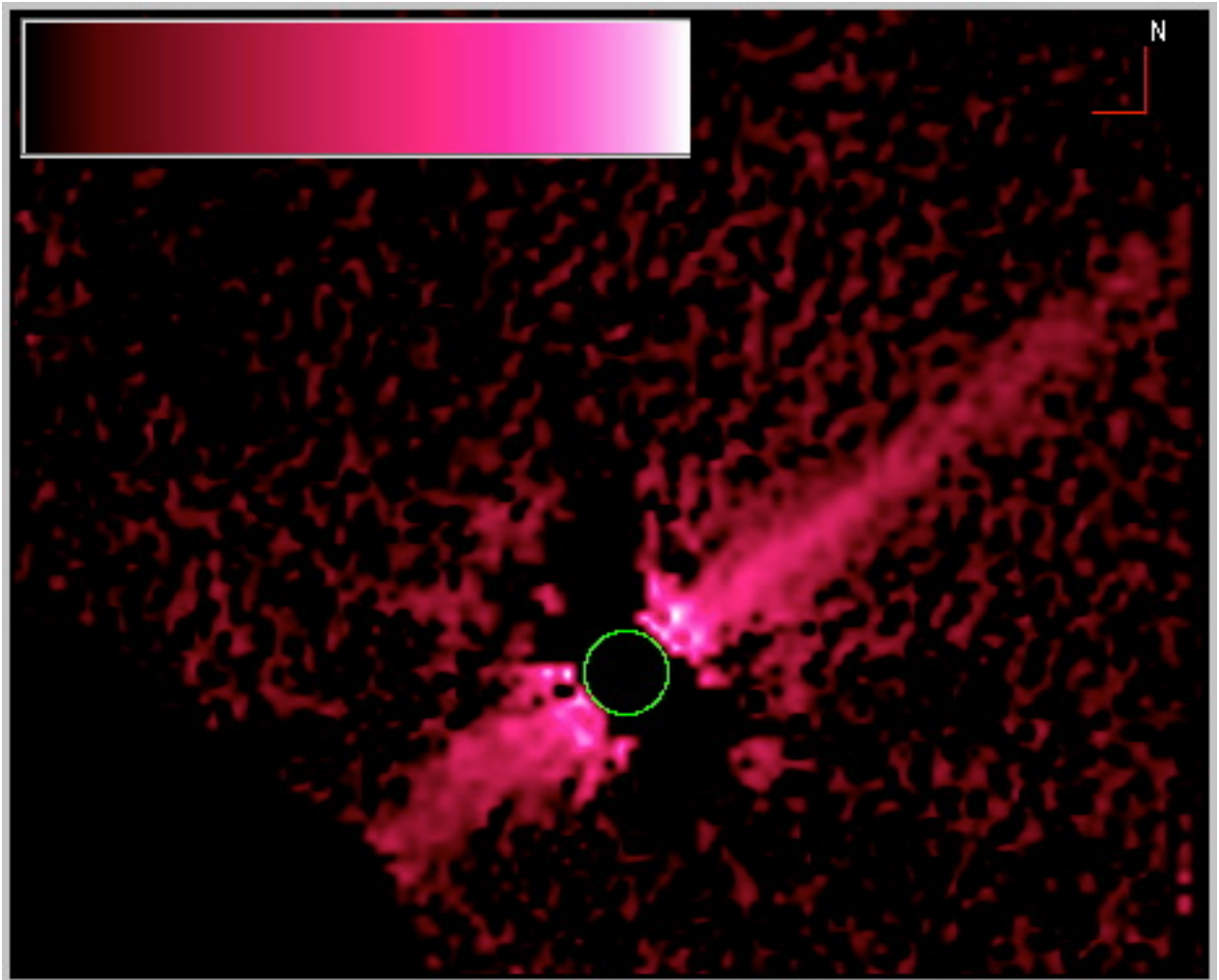

Fig. 7. Two-roll PSF-subtracted BAR5 imaging of the AU Mic inner disk Visit 04 & 06 best finger-centered images (3rd exposure set in each visit, planned for +0.08" offset). PSF template using star co-registered average of PSF template images from 2nd & 3rd exposure sets (planned for +0.04" and +0.08" offset). $\text{Log}_{10}$ image display dynamic range [-0.3] to [+2.0] {dex} counts $s^{-1}$ $pixel^{-1}$. The green circle is r = 0.20" (2 AU at AU Mic's distance of 10 pc). Total integration time combining all images used in both rolls is 23 s.

*10.3 Comparison to Prior Deeply-Exposed Wedge-A Results*

In *HST*/GO 12228 we obtained deep imaging of the AU Mic disk using 6 *HST* orbits at differing spacecraft rolls, coronagraphy at two different WedgeA occulting positions: WedgeA-1.0 and Wedge-A0.6, and utilizing a total of ~ 12 ksec of integration time. This produced a very high (photometric and astrometric) quality image for analysis, but with an achieved inner working angle limit of r = 0.5" (r ≈ 10 pixels). In *HST* GO/12923 (Fig. 7), using BAR5, we newly imaged the inner part of the larger FOV explored in GO/12228, 2.5x closer than had previously been possible. In Fig. 8 we reproduce the GO 12228 (Sch14) 6-roll combined image of the AU Mic "inner disk" region, in detail comprised of a 3.3 ksec integration using WedgeA-0.6 for the innermost part of the disk at r < 0.7"- 1.0", and an additional 9.8 ksec using WedgeA-1.0 beyond. We show this along with our 23 s total integration time GO/12923 two-roll combined BAR5 image (same data as in Fig. 7) at the same image scale, orientation, and display stretch. The image is derived from a single scan position with five 2.3 s exposures from Visits 04 and 06 of AU Mic. PSF subtraction was done using PSF templates from scan positions 4 & 5 combined from Visit 05. The GO 12932 image noise floor is, obviously, significantly higher – due to the differences in the observations, notably only 23 s of total integration time compared to up to 12 ksec, but also only two-rolls on the target compared to six. Despite the differences in sensitivity for these reasons, the IWA achieved in the bright, inner, region of the disk in GO/12923 improves over the GO/12228 imaging by a factor of 2.5. I. e., reducing the IWA from r = 0.5" (GO/12228) to r = 0.2" (GO/12932). The higher fidelity GO/12228 image serves to validate the disk photometry in the regions commonly sampled in the GO 12932 image (which agree very well, despite the "noise", as can be seen in these side-by-side images); thus indicative of no significant amount of stray starlight outside the BAR5 centered subarray in the commonly measured region r > 0.5" from the star.

---

[2] STIS employs a Lyot stop with outer radius 0.835 of the telescope pupil for an effective diameter of 2.00 m. Thus, at a pivot wavelength of ~ 0.58 μm, a diffraction limited resolution element (resel = 1.22 λ/D) is 72 mas = 1.44 pixels.

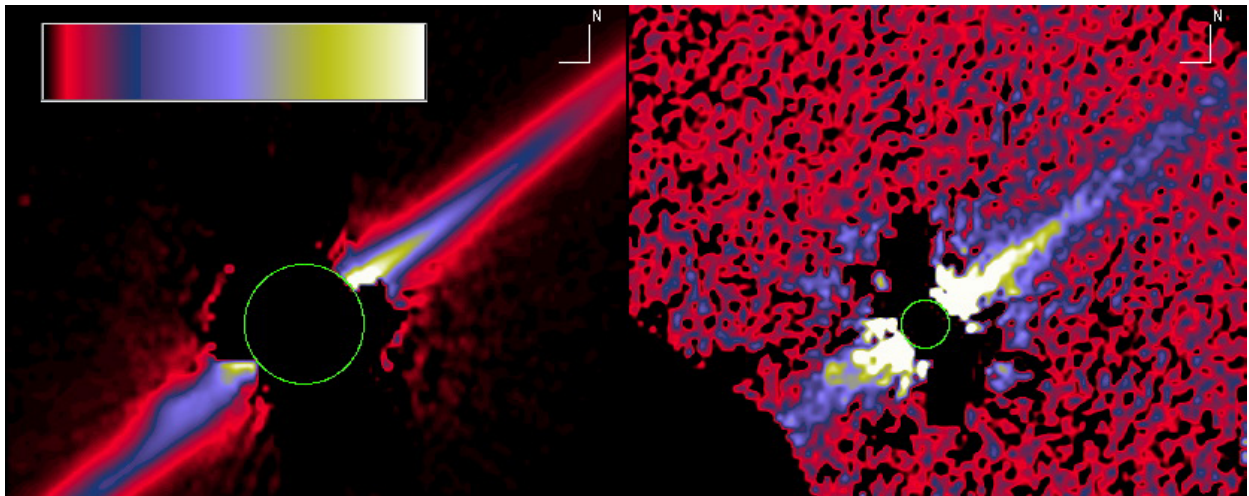

Fig. 8. GO 12228 (left; green circle r = 0.5") and 12923 (right; green circle r = 0.2") images of the AU Mic disk commonly sampled. Same image scale, orientation, and linear display dynamic range: 0 to + 15 counts s$^{-1}$ pixel$^{-1}$.

## 11. Follow-On Imaging Results – β Pictoris (Visits 01 - 03)

The first epoch AU Mic image scan observations provided the necessary occulter metrology information to compute *post priori* pointing corrections for subsequent imaging. The computed offsets[3] were applied in redefining the POS TARGs of the follow-on β Pic and δ Dor (PSF) visits.

The *a priori* known intrinsic repositioning uncertainty in the deployment of the coronagraphic optics by the STIS MSM (resulting in deviations from perfect target centering) is only ~ ±1/4 pixel.

---

[3] see Visits 01 - 03 of http://www.stsci.edu/hst/phase2-public/12923.pro

To test/confirm the efficacy of derived pointing corrections necessary to optimize coronagraphic centering, performance, and effective IWA, we thus conservatively planned image position offsets 3x larger at the extrema then the MSM redeployment uncertainty and with inter-point scan position steps finer than in the initial AU Mic orbits.

Specifically, for β Pic (Visits 01 and 03) and its PSF star (Visit 02) we executed 7-point step-and-dwell coronagraphic imaging scans with incremental offsets of +0.015" (0.3 pixels). For the BAR10 LL and LR corners the target was initially placed +0.225" along the respective 45° lines across the rounded edges. For BAR5, the cross-bar scans went from -0.045" to +0.045" across the mid-line of the occulter. "X marks the spot" stellar position determinations for the BAR10 scans confirmed the target positioning from the updated POS TARGS within an uncertainty of ~ 1/4 pixel. For BAR5 we then performed PSF-template-subtracted coronagraphy for all permutations of the disk and template star (multi-exposure median combined) images from all scan positions. With the pointing corrections derived from the prior AU Mic observations used to command the scan steps, the predicted "best" position on the mid-line BAR5 was at scan step 04, (a "half step" from the middle of the scan), and ± ~1 step for both β Pic and its PSF star δ Dor.

In Fig. 9 we show all possible combinations of PSF subtractions from Visit 03, stretched to simultaneously show: (a) the β Pic disk, (b) the BAR5 occulter and its ± edge gradients, and (c) the *HST* diffraction spike residuals. The best, and smallest IWA, images of the disk result when both the *HST* diffraction spikes and the BAR5 opposing light and dark edges are simultaneously nulled, anticipated in PSF subtracted images when both the target and template stars are positioned on the mid-line of BAR5. By visual inspection of Fig. 9, this occurs for β Pic position 04 (central position, as planned) and PSF template position 03 -- the dark green cell in Fig. 9. I. e., the template star at its planned position 04 was mis-positioned by (only) -1 scan step in SIAF Y w.r.t the mid-line of the bar. This also indicates that the system metrology is stable to ≤ ± 0.015" on greater than monthly timescales, as we determined the requisite POS TARGs from the AU Mic observations acquired 2 months prior to the β Pic observations.

In Fig. 9, below each image, are the post-facto determined PSF template image position shifts in SIAF X and Y pixels that were required to minimize the diffraction spike residuals by co-aligning the target and template stars independent of their possibly de-centered locations from the mid-line of BAR5. As can be seen, the co-registration correction needed to produce the best image was a template shift of (-0.03, +0.11) pixels, i.e., ~ half of the 1/4 pixel absolute positioning uncertainty. I. e., with perfect placement, the best image would have been in the central cell of the 7x7 mosaic of scan position images.

*Given the ~ 1/4 pixel relative pointing precision and MSM redeployment uncertainties, we recommend for highest fidelity PSF-subtracted imaging that all BAR5 observations be executed with a linear 3-point position "dither" of ~ (-1/4, 0, +1/4) pixel orthogonal to the bar with post-facto stellar co-registration prior to image combination for science analysis.*

Very similar results were found from the Visit 01 step-and-dwell image scans (identically executed, but at different celestial orientation). Fig. 10 illustrates, side-by-side at the display same stretch as Fig. 9, the best three PSF-subtracted images from the Visit 01 and Visit 03 in the SIAF frame (so the disk rotates about the occulted star from Visit 01 to 03), while the diffraction spike residuals and BAR5 occulter imprint remains rotationally invariant.

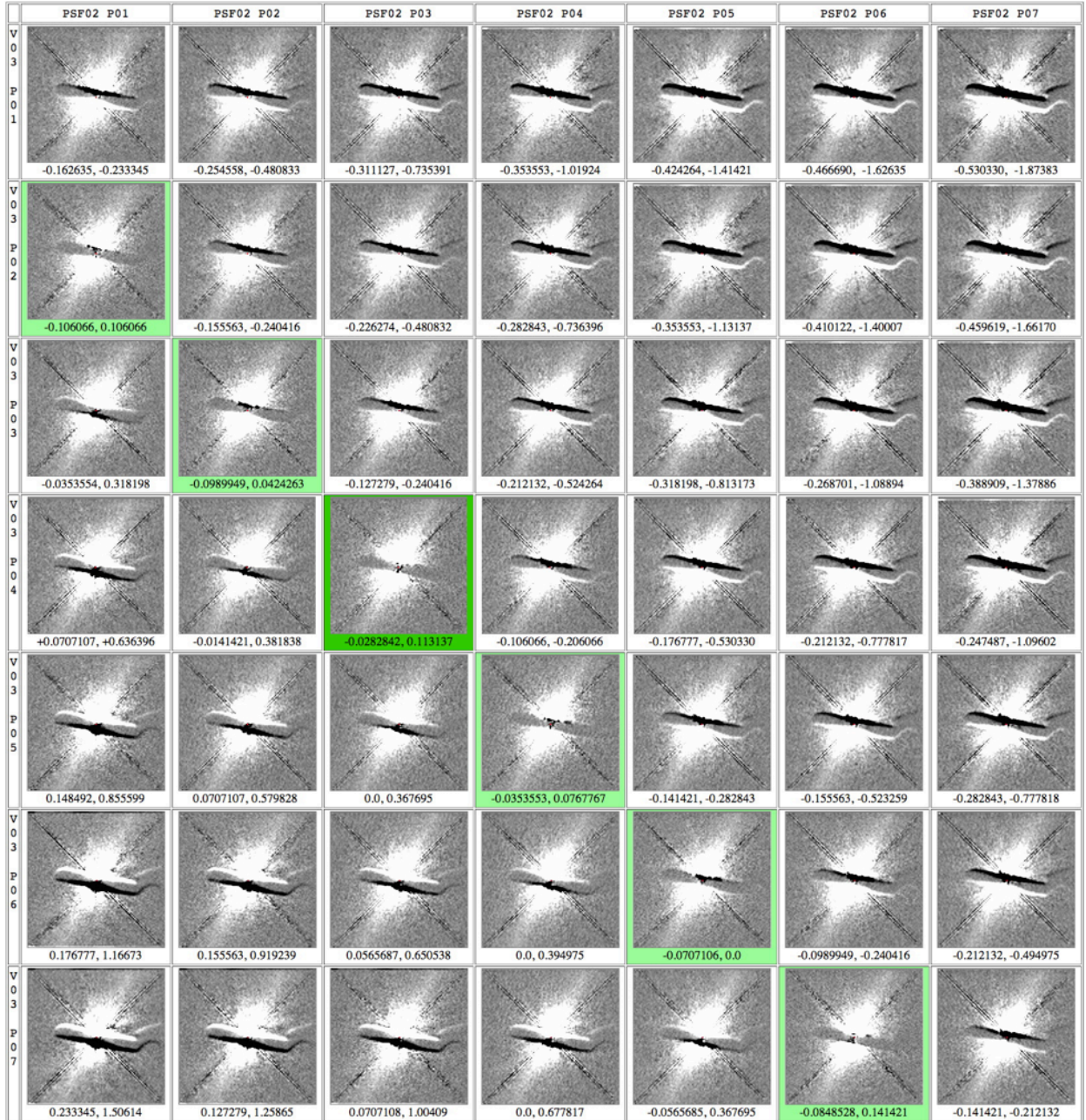

Fig. 9. Demonstration of the as planned coronagraphic target placements with BAR5, modulo ~ 1/4 pixel non-repeatability with MSM deployment of coronagraphic optics. All 49 Visit 03 PSF subtraction permutations are shown: β Pic (position 1 - 7, top to bottom) minus PSF template (position 1 - 7, left to right). The best image [PSF02 P03, V03 P04] (dark green cell) was anticipated at the central position [PSF02 P04, V03 P04] with both targets commanded to the mid-line position of BAR5, but was offset by -1 step in the template position due to ~ 1/4 pixel mis-centering of the image scan. Movement along the green diagonal preserves the 1 step in X and 1 step in Y relative positioning of both targets w.r.t. the BAR5 location, so bar edge brightness gradients remain shallow. In the orthogonal direction (lower-left to upper right) the co-aligned stars become increasingly decentered w.r.t. BAR5 with the opposing edge brightness gradients of opposite parity increasing toward the respective coroners.

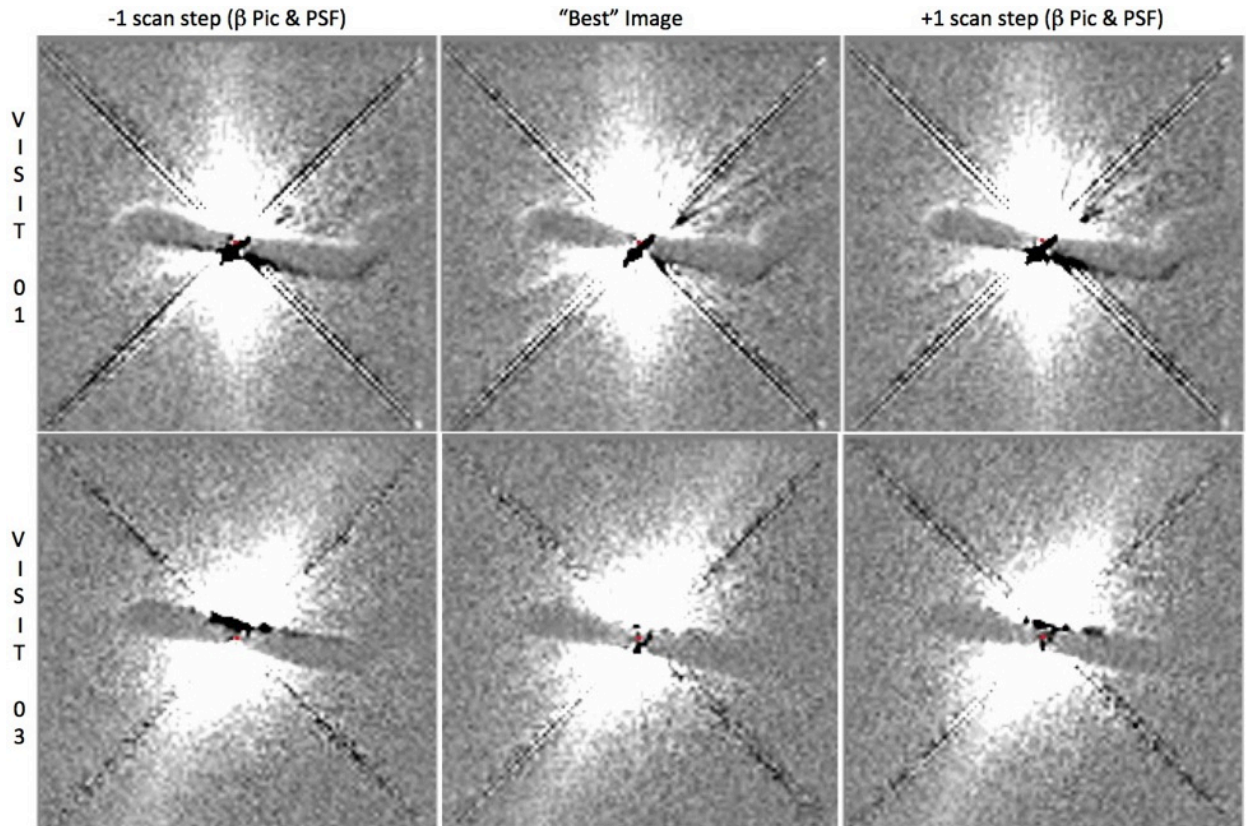

Fig 10. The as-determined best, and flanking, PSF-subtracted BAR5 images of the AU Mic disk at two celestial field orientations exploring as closely as possible the IWA space adjacent the BAR5 edges with small (1/4 pixel) cross-bar dithers, prior to later dither and roll combination into a single image. This is a demonstration observation using only two rolls. Enabled science observations using six field orientations would completely sample around the BAR5 occulter at all azimuth angles ≥ 0.15", and decorrelate quasi-static PSF-subtraction residuals in the celestial FOV (see Sch14 for analogous discussion and proof of concept with Wedge-A 0.6 and 1.0 occulting masks).

Prior to this enabling experiment for smallest IWA coronagraphy with *HST*, the β Pic disk had been most aggressively probed in GO program 12551 (Apai et al. 2015, ApJ 800 136) using a combination of: (a) STIS occulting wedges A and B at, (b) their 0.6" and 1.0" full-width taper positions, (c) with two spacecraft roll angles, (d) total integration time 1178 s, (e) contemporaneously interleaved PSF template observations. The reduced image of the inner part of the disk from the GO 12551 data set, combining all images in a north-up orientation while masking the "here be dragons" regions in the input images obscured by the STIS wedges and edge artifacts, is shown in Fig. 11 (left panel). The effective smallest IWA achieved, optimized in the GO 12551 observing plan[4] to image the edge-on disk in only two-rolls, was r = 0.35" – 0.40" (Apai et al. *ibid*) and larger at other stellocentric azimuth angles.

We compare the GO 12551 reduced image in Fig. 11 (left panel) to a similarly-combined image using the exploratory BAR5 data obtained in GO 12923 (Fig. 11, right panel) also at two rolls (Visits 01 and 03) but with total integration time only 3 s. The GO 12923 reduced image is derived from only the three cross-bar dithered images (in each visit) centered on and immediately flanking the BAR5 mid-line as shown in Fig. 10 in comparative accordance with our recommended 3-point dither imaging strategy. The BAR5 reduced image is derived from these six images. Though it is noisier than the GO 12551 image (see Fig. 11 caption), it successfully reveals the inner disk to an IWA ≈ 0.15" in the two-roll optimized direction along the edge-on disk plane after digitally masking the BAR5 occulter in the original images. In the regions commonly sampled, the morphology and surface brightness (SB) distribution of the β Pic starlight-scattering disk is well reproduced with the BAR5 imaging, with an ~ 2.5x improvement in IWA over the GO 12551 image. The high degree of morphological correlation in these images, gives confidence to the GO 12923 newly revealed image structure in the region interior to the GO 12551 IWA.

---

[4] see http://www.stsci.edu/hst/phase2-public/12551.pro for details

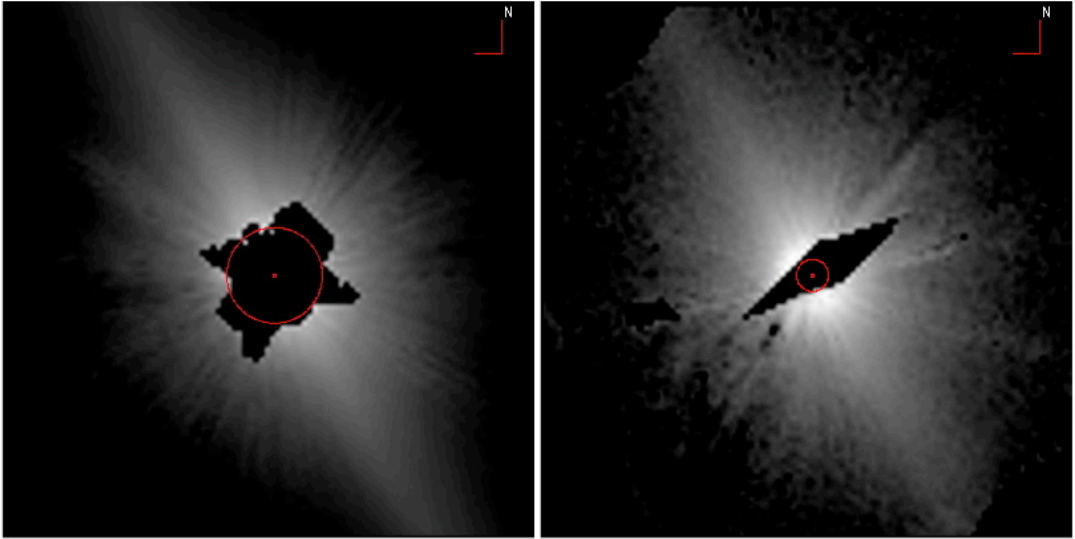

Fig. 11 compares the morphology, and verifies the imaging reproducability, of the inner β Pic edge-on disk at r ≲ 3" in the regions commonly sampled in the GO 12551 (left) and GO 12923 (right) data sets. The latter, designed to push the IWA, is limited in outer stellocentric distance by its short integration time and the subarray readout. These apples-to-apples image displays are optimized to allow a best comparison with the smaller FOV and more limited dynamic display range appropriate for the GO 12923 BAR5 data. The much more deeply exposed GO 12551 image, by virtue of its ~ 400x greater total integration time and use of four occulting wedge positions, has higher SNR and sensitivity to lower SB dust much further out (to r ~ 10") though is not illustrated in this figure; see Apai et al. 2015.

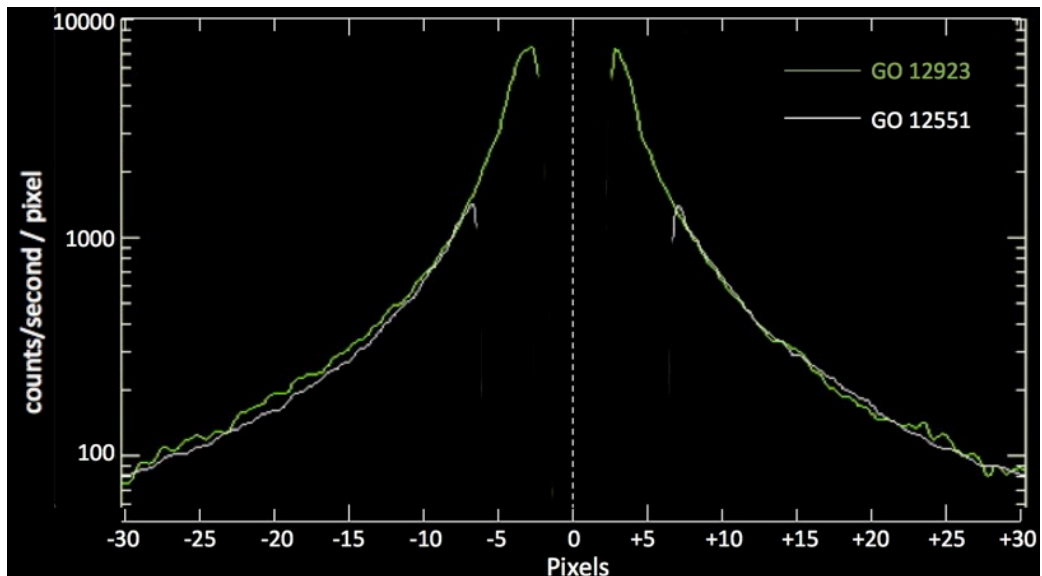

Fig. 12. Highly-correlated radial SB profiles (instrumental sensitivity in counts $s^{-1}$ pixel$^{-1}$ with CCGAIN = 4) of the β Pic disk in commonly sampled regions along the edge-on mid-plane. Measured to IWA limiting distances of ~ 0.15" (3 pixels) in the newly acquired GO 12923 image data and to ~ 0.35" (~ 7 pixels) in the GO 12551 data.

In Fig. 12, we overplot the radial SB profile of the edge-on β Pic disk measured identically from the independent GO 12551 (Wedge A/B, 0.6+1.0) and the GO 12923 (BAR5) reduced image data sets (i.e., from Fig. 11). The high degree of reproducability of the GO 12551 profile from the GO 12923 data in the regions commonly sampled validates the efficacy of the BAR5 disk photometry, and utility of the BAR5 occulter for narrow-angle PSF-subtracted coronagraphic photometry. The region interior to r ~ 7 pixels is measured only from the GO 12923 data, though the smoothly contiguous profile is indicative of the absence of image artifacts of significance beyond the masked edges of the BAR5 occulter.

## 12. BAR5 Starlight Suppression & Image Contrast

The ability to image faint point-, and low-SB spatially-resolved, sources in the close angular proximity to their much brighter host stars relies on observing techniques to suppress the background light due to the stellar PSF halo. With *HST*/STIS this is accomplished with a two-tier approach. (1) With coronagraphy using a simple Lyot stop and image plane occulters of different angular sizes. (2) Post-facto, through PSF-template subtraction further suppressing the stellar light in the PSF halo not rejected with coronagraphy alone. The image *contrast* at any unocculted field point provides a useful metric to assess the depth of the starlight rejection that then sets a detection floor for circumstellar objects. By contrast we mean: the ratio of the background light due to the coronagraphically suppressed PSF halo in a pixel (or resel) at any field point compared to the brightness of the central pixel (or resel) at the peak of the PSF if measured directly without coronagraphy or other methods of starlight suppression/rejection. Image contrast is both radially and azimuthally dependent, generally with improved (deeper) contrasts further from the star, but with azimuthally dependent structure at any radius. A contrast curve expresses the contrast as a function of stellocentric angle averaged, or medianed, in circumstellar annuli of increasing radii from the coronagraphically occulted star, i.e. an azimuthally averaged or medianed radial surface brightness profile in dimensionless contrast units.

To quantify the intrinsic starlight suppression with BAR5, we directly measured (in instrumental units of counts $s^{-1}$ $pixel^{-1}$) the SB of the coronagraphically well-centered δ Dor (Fig. 13, left panel) and HD 191849 template PSFs (Fig. 5, right panel) from the GO 12923 observations. In doing so we digitally masked the imprint of the occulting bar itself and the regions degraded by the unapodized diffraction spikes (e. g., Fig. 13, center).

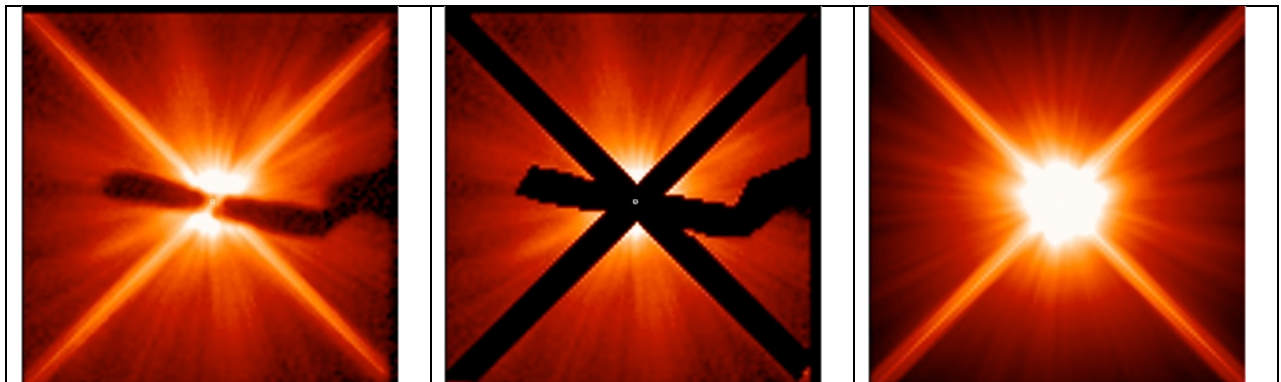

Fig. 13. Left: Image of BAR5 occulted PSF template of δ Dor from Visit 02, best-centered image scan position 4. Middle: Same as left panel with digital masking of the BAR5 occulter and OTA diffraction spikes. Right: TinyTim *unocculted* model PSF for an A7V star scaled to the brightness of δ Dor. All images $log_{10}$ display stretch from [+1] to [+4] dex counts $s^{-1}$ $pixel^{-1}$ with 101x101 pixel FOV centered on the star.

In the GO 12923 BAR5 enabling program we did not, however, image the unocculted stellar PSF template stars directly. Indeed, the BAR5 occulter was required to prevent image saturation at, and near, the PSF core. For background discussion on direct vs. coronagraphic imaging with STIS see Grady et al. 2003; PASP, 115, 1036. Thus, for direct to coronagraphic comparison (but not for subsequent PSF subtraction) we used TinyTim[5] model PSFs (Krist et al. 2011; Proc. SPIE, 81270J) in concert with the STIS imaging ETC (built on *synphot*) to obtain very close estimates for the flux (instrumental count rate) in the central pixel of the model PSFs: $2.63 \times 10^7$ counts s$^{-1}$ pixel$^{-1}$ for δ Dor, and $1.26 \times 10^6$ counts s$^{-1}$ pixel$^{-1}$ for HD 191849. These scale factors were used to transform the SB profiles in instrumental units of counts s$^{-1}$ pixel$^{-1}$ of the BAR5 images to a dimensionless contrast curves. Fig. 14 shows the raw BAR5 coronagraphic contrast curves as measured from the masked PSF images, calibrated through a direct image based on the central-pixel scaled TinyTim model PSF.

Despite their significant differences in brightness and [B-V] color, thanks to the instrumental stability, the azimuthally medianed contrast curves for δ Dor and HD 191849 are essentially identical. The greater dispersion about the medians for HD 191849, increasing with stellocentric angle, are due to the relatively shallow depth of the only 2x longer integration time (constrained by the experimental design) compared to δ Dor though HD 191847 is 3.6 magnitudes (28x in flux) fainter.

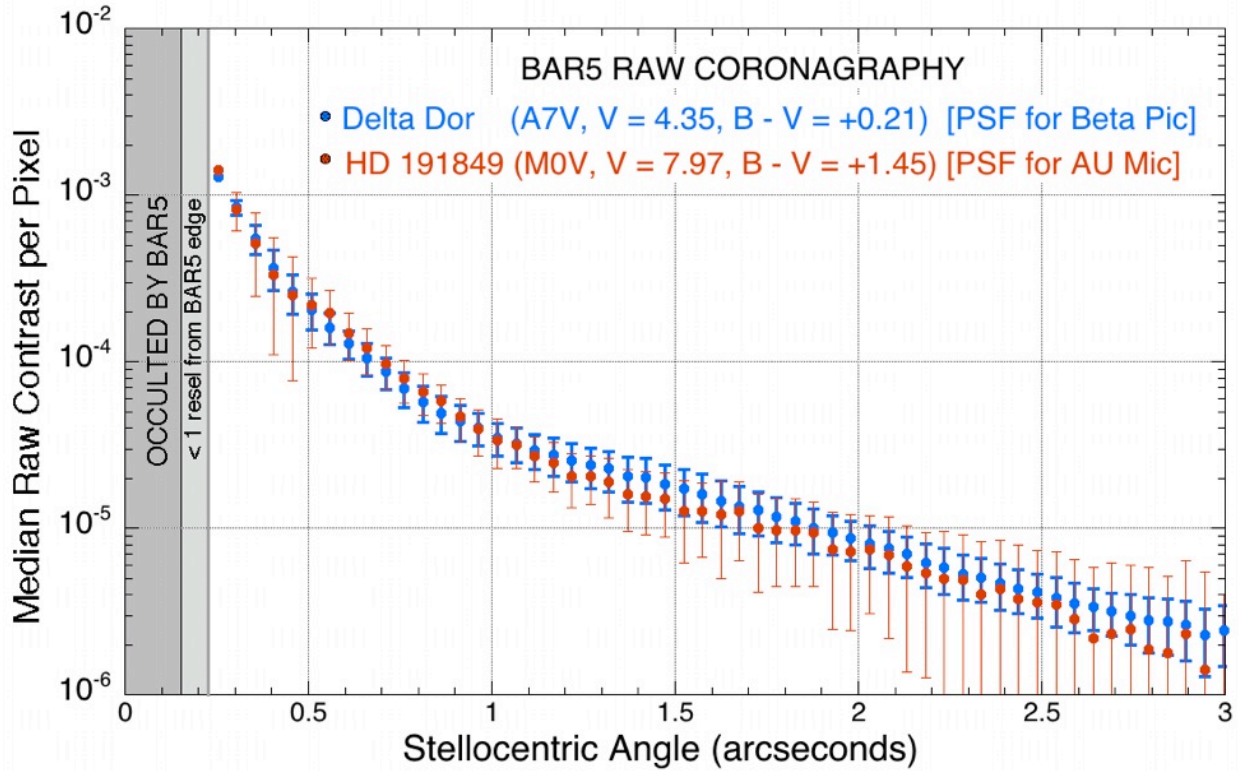

Fig. 14. BAR5 raw coronagraphic contrast (i.e., without PSF subtraction) – radial dependence. 360° azimuthal median with the BAR5 occulter and OTA diffraction spikes excluded (as shown in Fig. 13, middle panel). Error bars represent the 1-sigma dispersion in contrast around the median values. The angular distance from the star to the physical edge of the BAR5 occulter, and one full resolution element (~ 72 mas) beyond is indicated.

---

[5] http://tinytim.stsci.edu/sourcecode.php

## 13. BAR5 Contrast Augmentation with PSF-Template Subtraction

In GO 12923, the δ Dor PSF template was applied to the β Pic images to unveil the disk, but no calibrator for the template star itself was obtained – but is needed to show quantitatively the benefits of PSF subtraction clearly (i.e., in the absence of a disk signal). This was beyond the scope of the GO 12923 orbit allocation. To that end, however, here we make use of PSF calibration observations from the GO science program 13786 (Schneider et. al. 2016 AJ 152 64) that followed the commissioning of BAR5 and used the newly enabled occulter in the manner suggested in this ISR.

In GO 13786, HR 4735 (V=5.56, B9V, central pixel count rate 9.86 x$10^6$ counts s$^{-1}$ pixel$^{-1}$ with CCDGAIN=4) was used, as a contemporaneously imaged PSF template (applied to observations of the HR 4796A debris disk at two epochs six months apart). The two-epoch set of reduced BAR5 HR 4735 template images is shown in Fig. 15, left and middle panels. Importantly, they were obtained widely separated in time with very different Sun, beta, and roll angles, and hence with the OTA in different thermal states driving wavefront instabilities; *exactly* the systematics one would normally take effort to mitigate or minimize through observational design. Here, as a demonstration observation, we advantageously use the secular non-reproducibilities in these two different images of the same star in quantifying the raw repeatability, and PSF-subtracted image contrast limits.

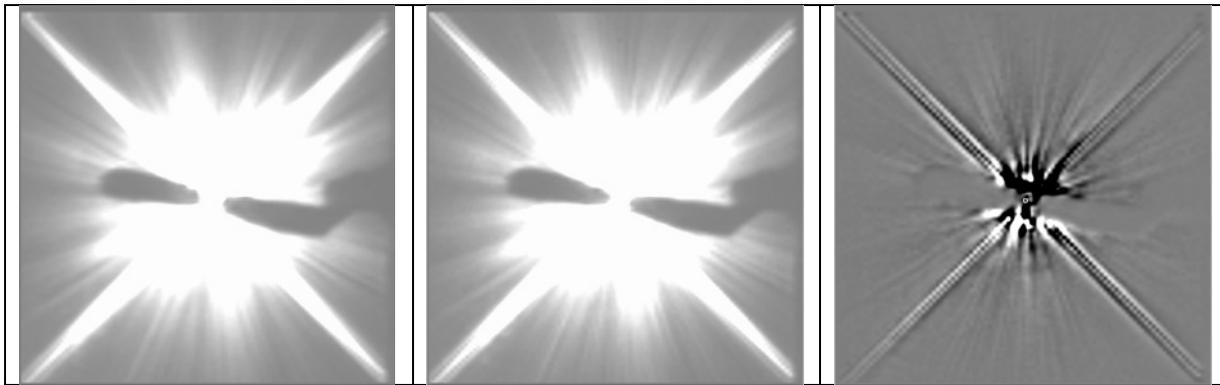

Fig. 15. Left: Well-centered BAR5 PSF template image of the star HR 4735 obtained in GO program 13786 Visit 43. Subtracting a second identically exposed and reduced image of the same star obtained six months later (from Visit 47 of the same program, middle panel), with the telescope in a different thermal pointing profile, reveals contrast-limiting differences in PSF structure (right) quantified in Fig. 16. Both the raw BAR5 coronagraphic, and PSF-template subtracted images, are shown with the same linear display stretch chosen to best show the PSF-subtraction residuals close to the BAR5 edges.

With a prior, fully independent determination of the BAR5 raw contrast curve as measured from the GO 12923 imaging of δ Dor (Fig. 14), in Fig. 16 we present raw contrast curves derived from the first and second epoch observations of HD 4735. Both of these curves, as in the case of δ Dor that is overplotted in this figure, are 360° azimuthal medians with the regions degraded by the OTA diffraction spikes, and obscured by the BAR5 occulter itself, excluded. Despite the low-amplitude differences in the fine (high spatial frequency) image structure revealed in Fig. 15 (right panel), the azimuthal median contrast curves for both HD 4735 template observations are virtually indistinguishable. Both, also, are nearly identical to the δ Dor curve that does deviate by a small amount only very close to the BAR5 edge, and (inconsequentially) beyond ~ 2". For clarity of presentation we did not overplot error bars in Fig. 16, though those found for both epochs of the HD 4735 contrast curves very closely replicate those found for δ Dor presented in Fig 14. I. e., all three raw contrast curves are very highly repeatable.

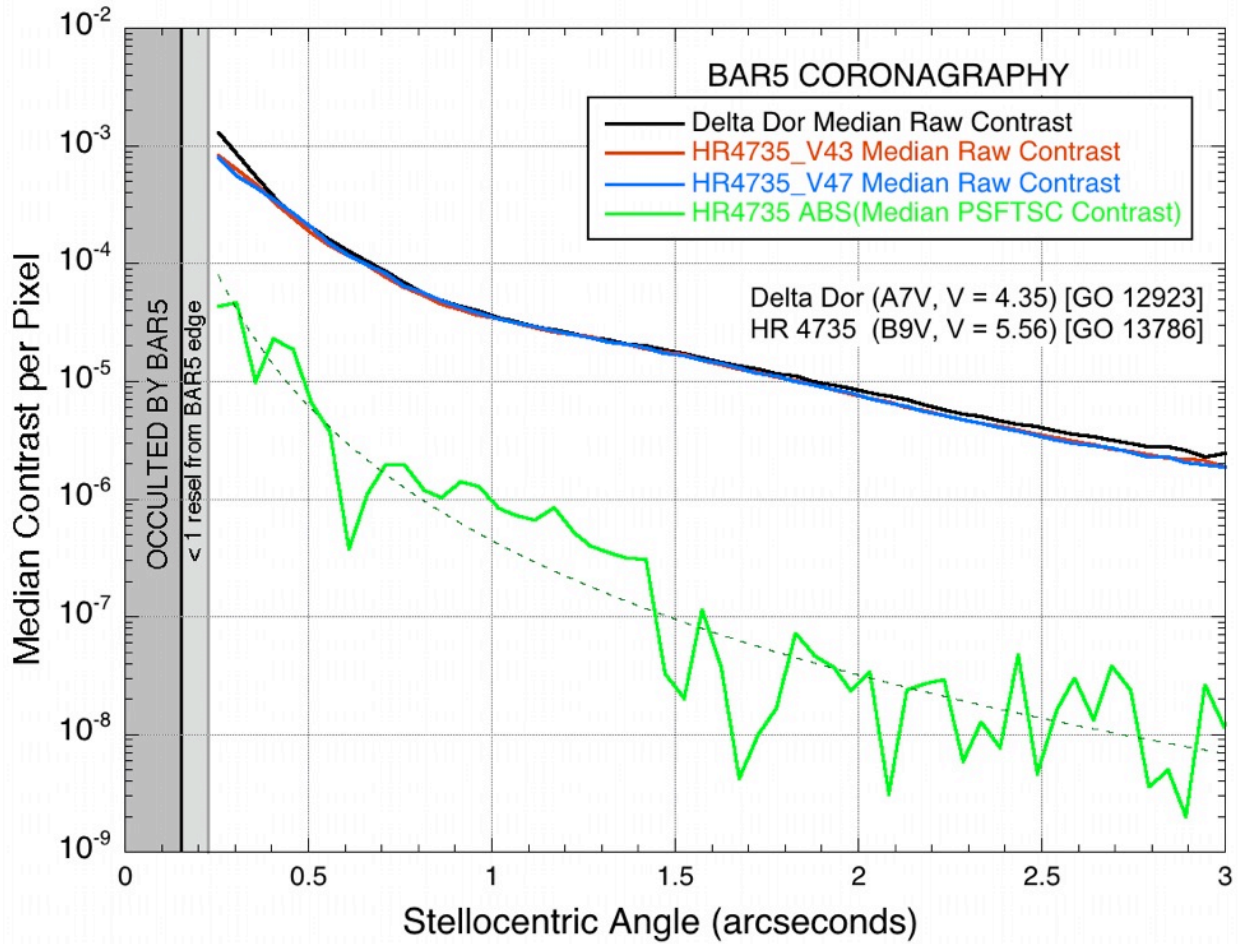

Fig 16. Raw (top; black, red, blue) and PSF-template subtracted (bottom, green) 360° (excluding diffraction spikes) median contrast curves for BAR5 coronagraphy. The raw contrast curve determined from observations of δ Dor (black), used as a PSF subtraction template for β Pic (with results presented earlier in this ISR), is compared to two highly-repeatable, independent contrast curves derived from similarly executed observations of HR 4735. The latter were then used to produce the azimuthal median BAR5 PSF template subtracted coronagraphy (PSFTSC) contrast curve (green).

To the eye, the two HD 4735 PSF template images in Fig. 15 (left and middle panels) look identical. They do, however, have differences in image structure, beyond photon noise, due to secular instabilities and non-repeatabilities in the telescope plus instrument system that are revealed with "self-subtraction" (Fig. 15, right panel). Such differences (PSF subtraction residuals) also appear imprinted (as "noise" or "image artifacts") on PSF-subtracted images of circumstellar disks and close-angular proximity faint companions. These residuals, however, are more are readily and cleanly seen using the star as its own (imperfect) PSF-subtraction template. In the non-photon noise dominated regime, these PSF-subtraction residuals ultimately set the contrast-limited detection sensitivity.

It is visually apparent, from inspection of Fig. 15, that PSF-template subtraction has done a laudable job in reducing the residual light in the circumstellar PSF halo. To quantify the level of starlight suppression achieved, we use the PSF-template "self-subtracted" image of HD 4735 as shown in Fig. 15 (right panel) to assess the with BAR5 coronagraphic contrast limits with PSF subtraction. This is shown in the green curve in Fig. 16. With PSF subtraction, the residuals can be (and are) biased both positively and negatively with respect to the true background. I.e., imperfect nulling will result in both positive or negative signals (e. g., light and dark compared to "true gray" in Fig. 15, right panel), of non-astronomical origin, setting the contrast floor. Thus, the azimuthal median residual amplitudes that set that floor are plotted in absolute value, i.e., the median absolute deviation[6] (MAD). We overplot the best power-law fit to the measured 1-MAD contrast limit (dashed line) that may be smoothly approximated as:

PSF-template subtracted 1-MAD contrast ≈ 4.45 x$10^{-7}$ x r $^{-3.80}$ ; with r in arcseconds

*A note on "color" and companion & disk imaging:* Using the same star as its own PSF subtraction template has the benefit of fully eliminating chromatic differences in PSF structure that are otherwise apparent with stars differing in optical color indices. Fully eliminating this one variable (difference in template "color") provides a result equivalent to ideal "color-matching" with a different template star. This (by elimination) decouples image residuals in PSF-subtractions due to mis-matches in template star colors from those that are more difficult to control except (for optimal science) through contemporaneous, and pointing-constrained, PSF calibration observations. As an observing technique, this two-roll method is particularly useful for imaging faint point-source companions where the roll angle difference is sufficient for the companion stellocentric distance to result in well separated positive and negative imprints of the companion in the difference image. This two-roll "self-subtraction" method, avoiding chromatic PSF-subtraction artifacts, also has utility for disk detection for edge-on and high inclination disks (e. g., Heap et al. 2000, ApJ 539 435; Krist et al. 2012, AJ 144 45). Augmentation with additional rolls and dither positions is discussed on: http://www.stsci.edu/hst/stis/strategies/pushing/coronagraphy_bar5. However, intermediate to low inclination (face-on) disks will suffer from partial to total flux cancellation with "self-subtraction". The mitigation of this effect requires, instead, PSF-template subtraction with a well color-matched PSF template star.

## 14. Using the BAR10 Rounded Corners

The STIS image plane coronagraphic mask includes a 10" long by 3" wide rectangular occulter with rounded edges extending into the FOV from the "top" (SIAF +Y) of the frame known as BAR10; see Fig. 1. The conceived utility of BAR10 was for wide-angle coronagraphic imaging of the circumstellar fields around the brightest stellar targets such as Fomalhaut. I.e., imaging low surface brightness CS dust or low-mass companions at large stellocentric angles from the brightest disk- or planet-hosting stars. BAR10 is wider than the tapered "tops" of either of the occulting wedges. BAR10 was designed to be used by placing a target in the middle of the bar, or possibly pushed "downward" (with a POS TARG) closer to its bottom edge, but still along the mid-line of the bar. Despite its posited utility for such observations, BAR10 has found very little use since it was commissioned in Cycle 7, with Wedge A/B occulters favored by observers.

The GO 12923 program was devised to enabling narrow(est)-angle coronagraphy with the 0.15" half-width BAR5 occulter (10x narrower than BAR10). Success, however, was not *a priori* assured. Hence, we conceived a possible, non-conventional, use of the BAR10 occulter for off-centered narrow-angle coronagraphy by placing a star near one (or both) of its rounded corners. Each would provide a FOV beyond its IWA covering ~ 270° in stellocentric azimuth angle. BAR10 was not designed for this purpose, and the fabrication quality (precision in manufacturing) for starlight rejection at and beyond its rounded edges at the extreme apices of the mask was unknown. We thus took the opportunity, in GO 12923 to explore this possibility.

---

[6] MAD = median( |$X_i$ - median($X$)| ) ; with a sufficient number of samples 1-MAD = 0.674-σ.

*14.1 BAR10 LL and LR Rounded-Corner Imaging Scans*

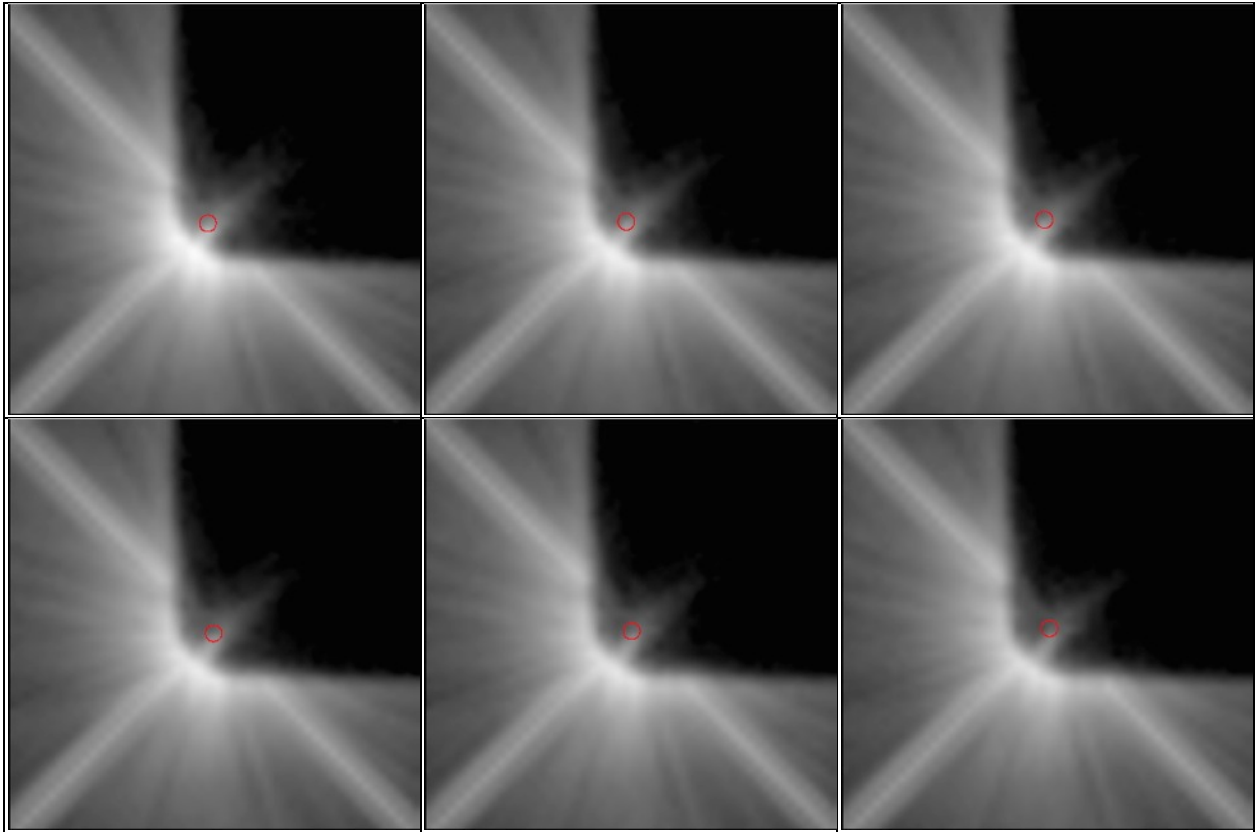

Fig 17. BAR10 LL rounded-corner image scans (β Pic visit #1) with incrementally rastered star location (red circle; drawn with r = 1 pixel [0.05077"] for scale) along the 45° diagonal in +0.015" commanded steps (top left to bottom right) showing the resulting rounded-edge (IWA) illumination. Star-centered FOV extract shown is 40 x 40 pixels.

    To explore the potential utility of BAR10 "rounded-corner" coronagraphy, we conducted step-and-dwell image scans using β Pic and its PSF template star δ Dor. This was done in the second set of visits (#'s 1 - 3) after executing and evaluating the earlier coarser scans (visit #'s 4 - 6) that were used to verify/update aperture positions and pointings with AU Mic as previously described. We found from those data that placing a star closer than ~ 0.225" from the BAR10 rounded corner edges would produce too much edge-illuminated light to be useful - see Figs. B-2 and B-3. We then performed finer-spaced BAR10 imaging edge-scans from 0.225" to 0.300" (the half-width of WedgeA/B-0.6), repositioning the target in incremental steps of +0.015" interior to the occulter LL and LR rounded edge on a 45° diagonals (e.g., see Fig. 17). Fig. 17 illustrates the raw coronagraphic imaging results for the first β Pic visit[7] #1 (BAR10 LL corner). The presence, but incrementally decreasing level, of un-rejected edge-scattered light at and beyond the rounded-edge corner is visually apparent from the first to the last target position. Table 2 gives the actual stellar locations at each LL scan position (in SIAF coordinates), recovered from these data with "X-marks the spot" centroiding, used in the later generation of stellocentric contrast curves for each image.

---

[7] The β Pic disk is much fainter than the raw coronagraphic PSF halo and thus remains invisible without PSF-subtraction.

Preprint: Under Review              Instrument Science Report STIS 2017-## Page 21

| TABLE 2 -- VISIT 01 LL SCANS | | | Stellar SIAF Locations (in Fig. 17) | |
|---|---|---|---|---|
| data ID | Step# | Δ edge" | X | Y |
| OBZE01010 | 01 | 0.225 | 600.322 | 50.447 |
| OBZE01020 | 02 | 0.240 | 600.554 | 50.651 |
| OBZE01030 | 03 | 0.255 | 600.759 | 50.884 |
| OBZE01040 | 04 | 0.270 | 601.030 | 51.056 |
| OBZE01050 | 05 | 0.280 | 601.199 | 51.324 |
| OBZE01060 | 06 | 0.380 | 601.368 | 51.654 |

In Fig. 18 we present ~ 180°-azimuthally medianed radial surface brightness profiles for each of the six target positions shown as contrast curves. For contrast calibration, we adopted an estimate of the flux in a modeled unocculted central pixel for a β Pic PSF as 6.49 x$10^7$ counts s$^{-1}$ pixel$^{-1}$ by the same method described prior for δ Dor and HD 191849 in § 11. In stellocentric azimuthal medianing, we digitally excluded nearly all of the BAR10 mask area, except very near and including the rounded corner that is the source of potentially problematic light limiting the image contrast in this region of interest. The digital masks also excluded the "upper-left" to "lower-right" diffraction spikes, but did not exclude the "lower-left" spike that is superimposed upon the rounded-edge scattered starlight, as it would appear in the region of particular interest at small stellocentric angles; see the masked-image inserts provided in Fig 18. for illustrative examples of digital masking for scan positions 1 and 6.

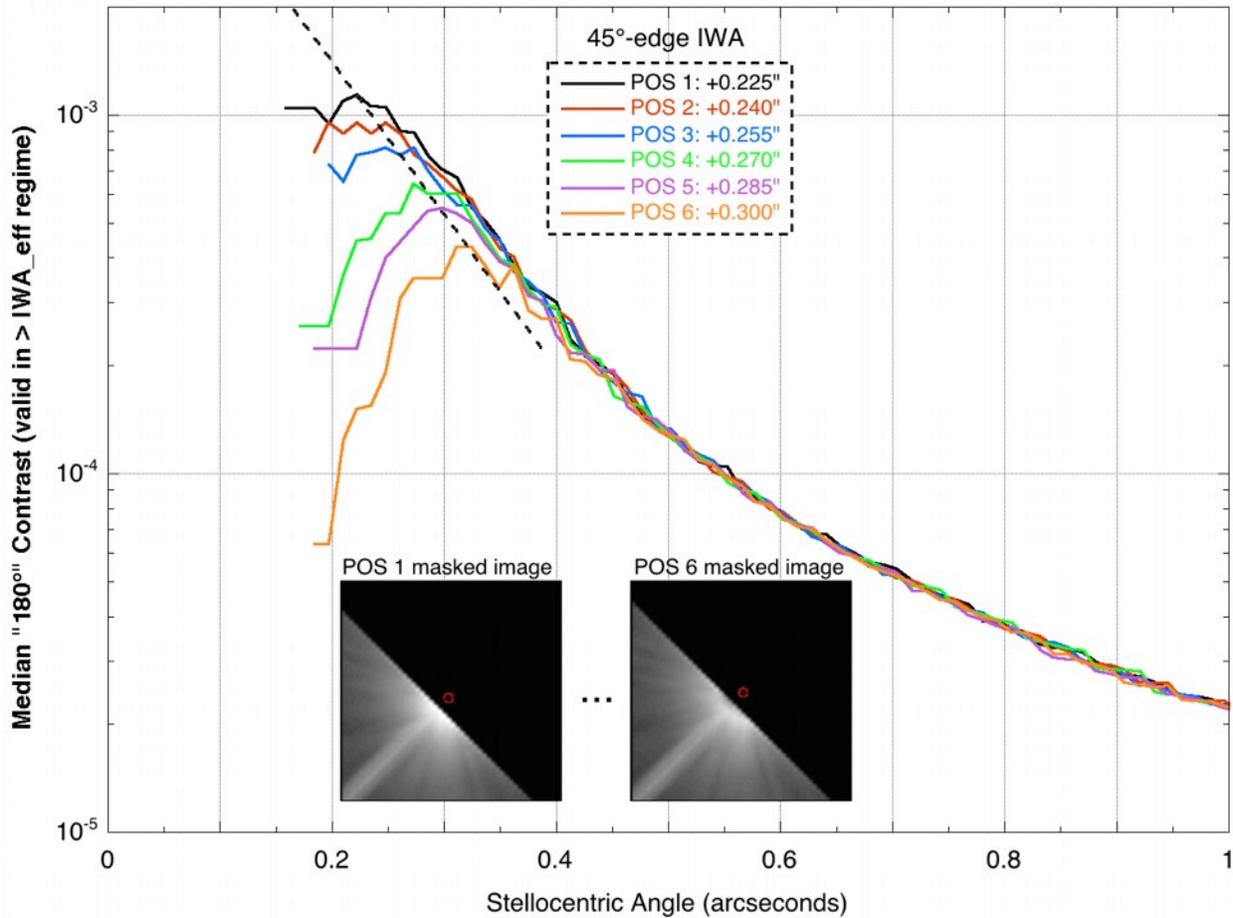

Fig 18. BAR10 LL rounded-corner contrast curves with incremental repositioning of the occulted star w.r.t. the rounded edge of the focal plane occulter along a 45° diagonal passing through the corner's center of curvature. Inset images illustrate the digital masks used when computing azimuthal median contrast excluding the UL-to-LR diffraction spikes (that move with the star when repositioned) and the BAR10 occulter itself; see Fig. 17 for corresponding unmasked images.

Each contrast curve turns over very close to the occulter-edge effective IWA, corresponding the distance from the star to the rounded-edge of the BAR10 corner. In Fig. 18 the locus of the turn-over points is connected, in approximation, by the black dotted line. To the left of that line the "turned-over" data are corrupted or invalid being impacted by the (small) digitally unmasked region of the BAR10 edge. To the right of this line, the data are valid and there the raw IWA vs. contrast performance may be assessed. As can be seen in Fig. 18, on the azimuthal median the IWA contrast is best (smallest) by a factor of ~ 3 with the star positioned +0.300" (POSition 6) compared to a smaller positioned IWA of +0.225" (POSition 1). I.e., at 0.300" the contrast is improved, but IWA comparatively suffers. The corresponding contrasts, however, suggest that with augmenting efficacious PSF subtraction, the BAR10 rounded-corners may provide additional useful capability.

The differences in the azimuthally medianed contrast performance diminishes at larger stellocentric angles, and vanishes beyond r ≈ 0.5"; see Fig. 18. Very similar results are obtained with the BAR10 lower right corner, and also with the second β Pic visit #3 and also for the δ Dor PSF visit #2. The two-dimensional morphology of this stellar background light in this region, however, is complex, as can be seen for the LL corner in Fig. 17. The BAR10 rounded-edge brightness pattern has dependency upon the two-dimensional (im)precision of the target centering w.r.t. the round-edge center of curvature location. This presents additional complexities, and uncertainties, in PSF-template subtraction that are obviated (in one dimension) with BAR5 straight-edge coronagraphy.

*14.2 BAR10 Rounded-Edge PSF Subtraction*

We next evaluated the efficacy of BAR10 rounded-corner PSF-subtracted coronagraphy. As is not atypical for STIS coronagraphy in general, the Visit 2 PSF images at same commanded scan-point offset positions as the β Pic Visits 1 and 3 images have (*a*) mis-alignments in the projected location of the BAR10 occulter on the CCD due to MSM repositioning non-repeatabilies. They also have (*b*) mis-alignments of the target placement w.r.t. the BAR10 itself (and as seen re-imaged at the CCD focal plane) from target acquisition imprecision that must be "corrected" in post-processing prior to PSF subtraction. (See Sch14 for details as discussed analogously for the Wedge-A occulter). (*b*) is correctable (except very close to the BAR10 edges) in post-processing by PSF image registration for (*a*) best-matched scan positions that approximately compensate for MSM position re-deployment non-repeatabilities. E.g., in normal science observations, as recommended separately for BAR5, with ± 1/4 pixel "dither" steps providing a full-range of a half-pixel offset at the extrema for target:template position matching and correction. The β Pic/δ Dor BAR10 exploratory image scans in GO 12923 used an inter-scan step size of 0.015" that is very close to 1/4-pixel.

Because of (*a*), in Visit 1 for β Pic at the LL corner of BAR10, only images at scan steps 3 – 6 could be matched to the Visit 2 LL images of δ Dor (at scan positions 1 - 4, respectively). I.e., there was an approximately two step mis-alignment offset in the BAR10 occulter imprint on the detector between the β Pic visit 1 and δ Dor visit 2. Each of theses four scan-position matched β Pic/δ Dor image pairs were then co-registered on the location of the occulted star, and the template images were subtracted with a flux scale factor of 1.55. The resulting set of PSF-subtracted images is shown in Fig. 19.

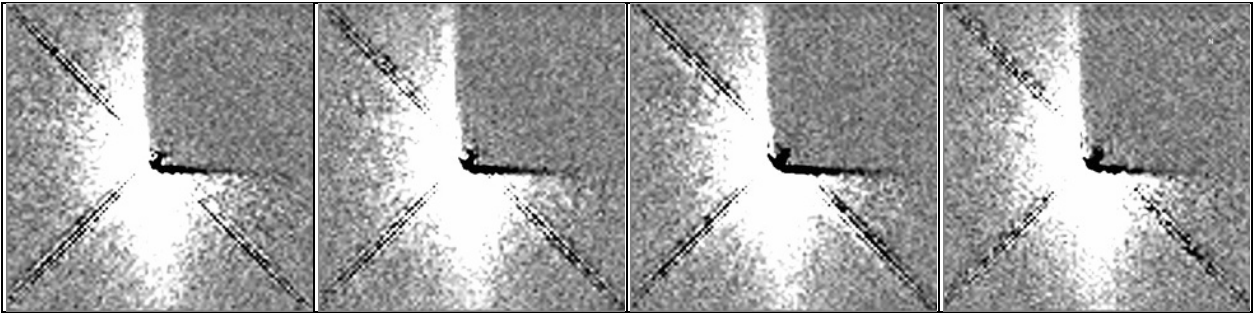

Fig. 19. PSF-template subtracted images of the β Pic disk with the star occulted by the BAR10 LL rounded corner. Left-to-right with incremental offsets of 0.015 arcseconds closely compensated in position matching of the BAR10 imprint in PSF image scans with the post-factor co-alignment (co-registration) of target and template stars before subtraction. 100x100 pixels (~ 5.1" square) FOV centered on the star.

Within each visit we separately executed identically designed, but "mirror symmetric", imaging scans at both the LL and LR BAR10 corners. Following the PSF-subtraction process described above, Fig. 20 (left and middle panels) illustrates by example the high degree of imaging repeatability in commonly sampled regions (i. e., "below" BAR10), and averaged in the right panel. Here we compare the (third) scan position in visit 1 at the LL (same image as left panel in Fig. 19) and LR corners.

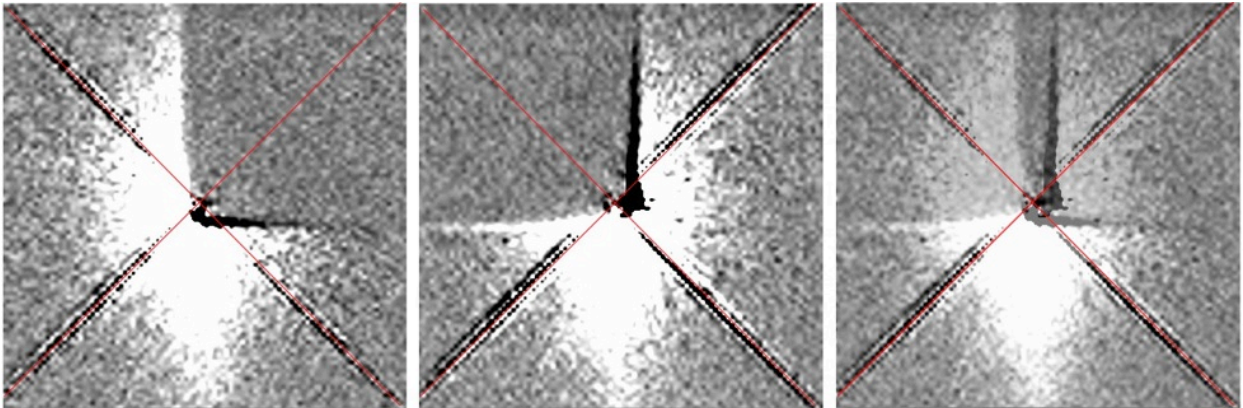

Fig. 20. PSF-subtracted images of the β Pic disk at the BAR10 LL and LR corners. Red X's mark the locations of the occulted star. Right panel, co-registered and averaged, illustrates how the coronagraphic fields overlap about the star.

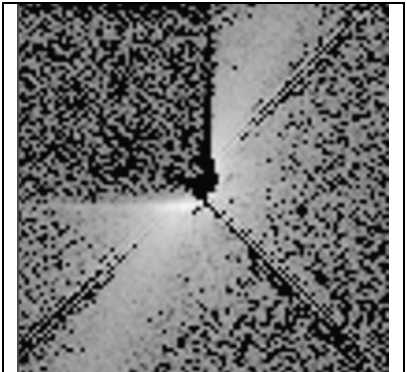

For either the LL or LR corners, BAR10 imaging provides an unocculted stellocentric field approaching 270° in azimuth beyond the effective IWA. There is a "gap" however in the overlapped regions set by both the IWA and artifacts along the BAR10 edges; see Fig. 20, right panel. Additionally, the post-facto stellar registration process results in BAR10 artifacts that are manifested as steep positive or negative brightness gradients that externally flank the bar edges. This can invalidate the data up to a few pixels beyond the physical edges. In Figs. 19 and 20, as presented, these artifacts appear as hard black (negative) or hard white (positive) near-linear features at the bar edges, through the latter (positive) is difficult to distinguish from the disk itself in the display stretch used for other illustrative purposes. In Fig. 21 we illustrate the typical extent of these edge artifacts in a stretch where more easily seen.

Fig. 21. Necessary co-registration of target and template PSF causes mis-alignment of the BAR10 profile resulting in gradient artifacts close to its edges.

To test the robustness of BAR10 corner PSF-subtracted coronagraphy in concert with a "standard" two-roll observing paradigm, we followed an identical process for the same β Pic scan point images (3-6) in visit 3. In this visit, the disk was re-oriented +30° w.r.t. visit 1, thus requiring independent MSM deployments and target acquisitions. In this case, best position matches for each of these scan points were found from the δ Dor LR corner scan positions 3 − 6 (i. e., the same as for the β Pic points, rather than 1 − 4 as in visit 01), respectively. Those resulting images are shown in Fig. 22. As can be seen, despite the differences in the scan-position dependent raw contrast floor close to the BAR10 rounded corner edges the images all exhibit a very high degree of repeatability.

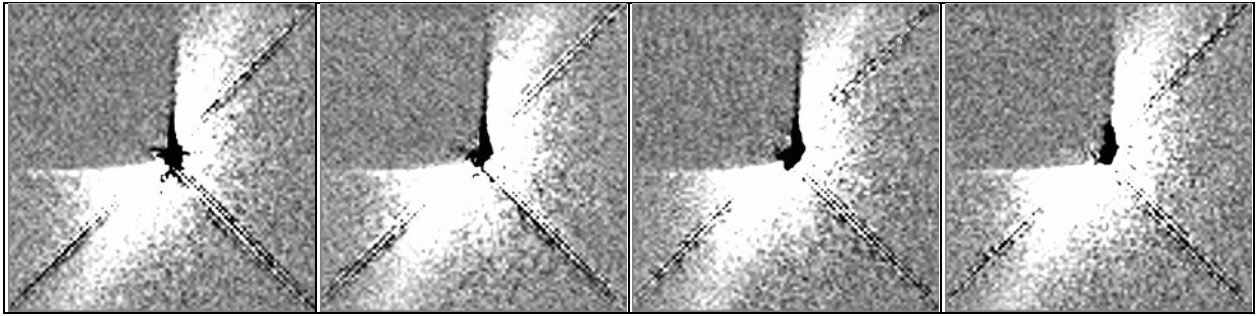

Fig. 22. PSF-template subtracted images of the β Pic disk with the star occulted by the BAR10 LR rounded corner, from visit 03 image scans (positions 3 - 6) with the disk axis rotated +30° clock-wise w.r.t. visit 01 (Fig. 19).

*14.2 BAR10 Rounded-Edge 2-Roll PSF-Subtracted Image Combination*

The rotation of the field by 30° from visits 1 to 3, in combination, reduces the gap in spatial coverage flanking the BAR10 vertical edges to a small triangular wedge beyond the IWA, as shown in Fig. 23. Fig. 23 combines the four LL and four LR PSF-subtracted images shown in Figs. 19 and 22 by co-registering the eight images at the location of the occulted star, rotating all to a common celestial frame, and individually applying digital masks to the BAR10 occulted, and (most) diffraction-spike affected, regions prior to median combination. The disk is recovered at r ≥ 0.36" along the NE major axis, and at r ≥ 0.46" along the SW major axis (being impacted interior by the remaining (only) two-roll gap in the LL + LR spatial coverage). The red circle indicates a stellocentric angular radius of 0.25", the smallest effective IWA over the range of azimuth angles where the disk had been recovered in the two-roll reduced image.

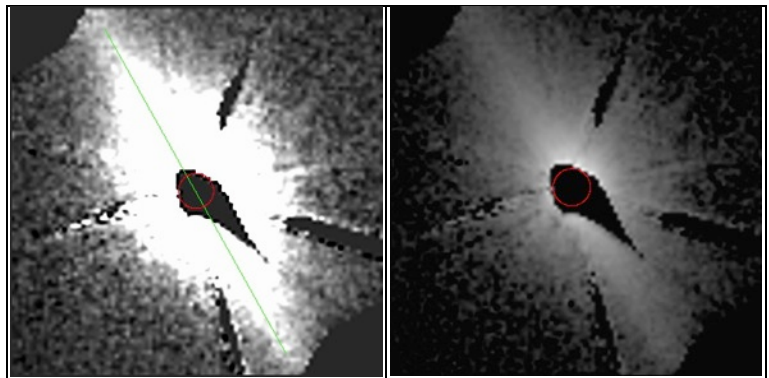

Fig. 23. Two-roll BAR10 LL and LR combined image of the β Pic circumstellar disk to a field limit of 100 x 100 pixels (5.1" x 5.1") commensurate with the total depth of integration of 0.8s for all eight images. Left: linear display from -20 to 100 counts $s^{-1}$ pixel$^{-1}$. Right: Log$_{10}$ display from [+1] to [+4] dex counts $s^{-1}$ pixel$^{-1}$. Both "north up" with the *a priori* known orientation of the disk major axis (PA = +29.1°) indicated by the green line in the left panel. Red circle: r = 0.25".

The morphology of the inner (r < 2.6") region of the β Pic disk is well recovered in the 2-roll BAR10 LL + LR combined image. To evaluate the photometric efficacy, we measured and compare in Fig. 24 the disk surface brightness profile along its major axis from the 2-roll combined image, to the *a priori* established profile previously ascertained using the Wedge A+B occulters in GO 12551, and as reported and compared in § 11 of this ISR with BAR5. Despite the very small number (8) of only short (0.1s) exposure time images contributing to the 2-roll combination used for comparison, the major axis radial surface brightness profile beyond the NE and SW effective IWAs, is well reproduced with a precision of a few percent. The BAR10 profile is systematically lower (by a few percent) on the NE (left) side of the disk. This may be due to the residual offsets in target:template positioning w.r.t. the BAR10 corners not fully remediated with quantized scan position matching.

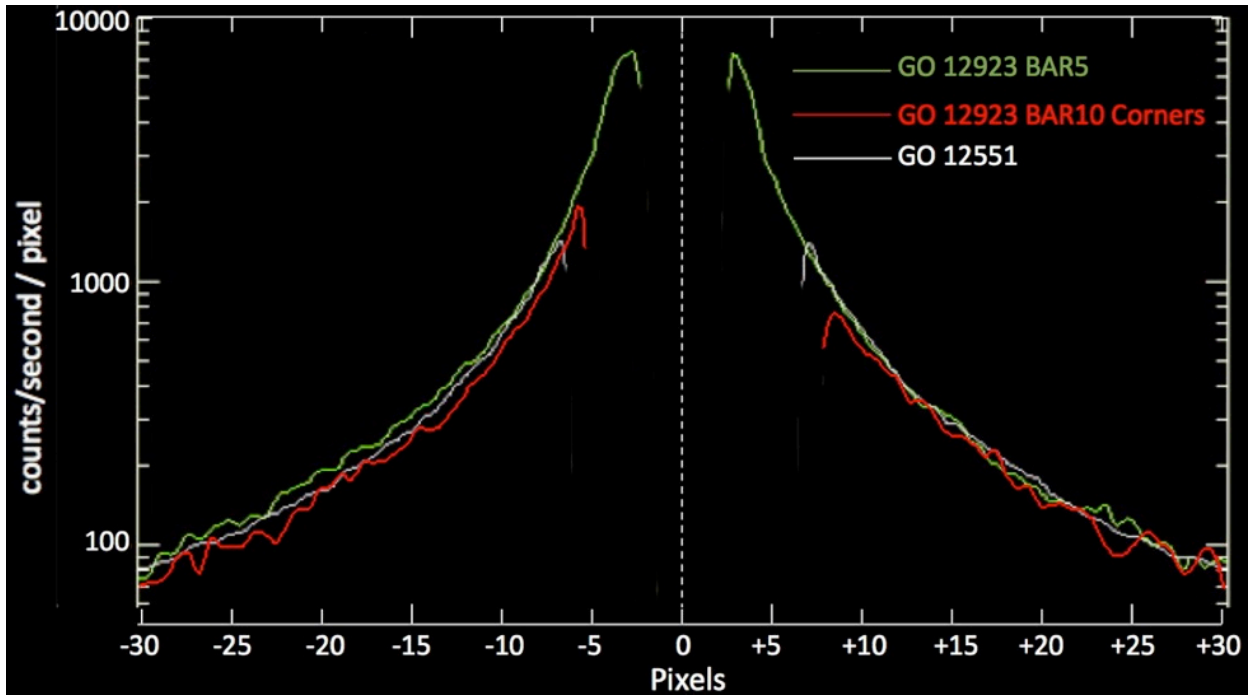

Fig. 24. Radial SB profile (in instrumental units) along the β Pic disk major axis comparing photometry from 2-roll combined PSF-subtracted images with BAR5, BAR 10 LL+LR, and (from GO program 12551) Wedges A+B/0.6+1.0.

### *14.3 BAR10 Corner Rounded-POS TARGs*

For the BAR5 finger, a target is always optimally placed when located on the mid-line of the bar exactly between the two, parallel, edges that are 0.3" apart. In § 15 of this ISR we give specific POS TARG offsets (from the BAR10 aperture location fiducial) to achieve such a pointing, specifically mid-way along the unbent (but tilted) long axis of BAR5.

For each of the BAR10 corners, in placing the target with respect to only a single rounded edge, the observer has a degree of freedom in placing the "back-off" distance from the edge along a 45° diagonal passing through the corner center of curvature based on a trade between inner working angle and contrast at r < 0.5" (see Fig. 18). Here we similarly give specific POS TARGs for the BAR10 LL and LR corners for effective IWAs predicated on the contrast "turn over" as shown in Fig. 18. Because of target placement imprecision, we recommend obtaining observations of both the target and PSF template star with a three-point dithers of 1/4-pixel steps centered on the desired POS TARG along the 45° passing through the rounded corner.

Table 3. POSition TARGet Offsets from BAR10 Reference Position for LL & LR Corner Imaging

|  |  | BAR10 Lower Left Corner | | BAR10 Lower Right Corner | |
| --- | --- | --- | --- | --- | --- |
| IWA | Raw Contrast | X Pos Targ | Y Pos Targ | X Pos Targ | Y Pos Targ |
| 0.225" | $1 \times 10^{-3}$ | -1.23270 | -1.30535 | +1.33805 | -1.30016 |
| 0.240" | $9 \times 10^{-4}$ | -1.22209 | -1.29474 | +1.32744 | -1.28955 |
| 0.255" | $8 \times 10^{-4}$ | -1.21148 | -1.28414 | +1.31683 | -1.27894 |
| 0.270" | $6 \times 10^{-4}$ | -1.20088 | -1.27353 | +1.30623 | -1.26834 |
| 0.285" | $5 \times 10^{-4}$ | -1.19027 | -1.26292 | +1.29562 | -1.25773 |
| 0.300" | $4 \times 10^{-4}$ | -1.17966 | -1.25232 | +1.28501 | -1.24712 |

## 15. Summary and Recommendations

1) Use of the STIS BAR5 coronagraphic occulter, though damaged pre-launch, has been verified and validated as functional to produce high-contrast images in close ($\geq 0.15$") angular proximity to bright point sources, and for the BAR10 LL and LR "rounded corners" with targetable IWAs in the range $\geq 0.225" - 0.300"$.

2) Target pointing offsets and/or aperture metrology updates as codified in the PDB SIAF.dat file and/or implemented through APT are required to properly position a target centrally on the mid-line of the BAR5 occulter, or with respect to the BAR10 LL or LR corners, to achieve useful/optimum coronagraphic starlight suppression.

3) Item (2) pointing corrections may be implemented in the STScI/*HST* ground system either as POS TARGs from the existing BAR10 aperture, or by the equivalent update/definition of a BAR5 aperture position (and/or BAR10LL/BAR10LR mnemonics) in SIAF.dat and flowed down to other elements of the ground system.

*N.B: We specify (2) throughout this ISR as POS TARG offsets from the BAR10 aperture pointing fiducial as propagated into the STScI ground system observation planning S/W systems per the PDB SIAF.dat file circa mid-2013. STScI may elect to update the BAR5 fiducial and/or develop new aperture definitions for the BAR10 LL and LR corners based upon these offsets. If so, observers should be aware not to then additionally apply those offsets if built into updated aperture definitions.*

4) Optimum mid-BAR5 pointing (with uncertainty of $\pm \sim 1/4$ pixel) is achieved with:

   POSTARG (17.48579, -7.42153) from the BAR10 aperture

5) Because of the relative narrowness (0.15" half-width) of the BAR5 occulter, and given the *a priori* known non-repeatabilities of the STIS MSM that deploys its coronagraphic optics, all BAR5 observations (target and PSF template) would greatly benefit from being executed as a 3-point linear cross-BAR5 "dither" with the central point at a planned mid-BAR position, and flanking positions at +/- 1/4 pixel orthogonal to the bar. Such observations are implementable, simply, with the following Phase 2 special pointing requirements:

   POSTARG 17.48331, -7.43398 from BAR10 aperture; 1/4 pixel below nominal BAR5 mid-line
   POSTARG 17.48579, -7.42153 from BAR10 aperture; on nominal mid-line of BAR5
   POSTARG 17.48827, -7.40908 from BAR10 aperture; 1/4 pixel above nominal BAR5 mid-line

Similar 3-point imaging scans should be conducted for BAR10 LL and LR corners about a chosen IWA location (Table 18) on a SIAF 45° diagonal (passing through the nominal center of curvature).

6) BAR5 observations have been demonstrated, for both "red" (B-V > 1) and "blue" (B-V ~ 0) targets, with an ability to image, resolve, and recover astrophysical circumstellar structure at stellocentric angles as close as 0.15" – 0.20" from an occulted point-source.

7) Instrumentally calibrated PSF-subtracted BAR5 and BAR 10 LL/LR corner photometry has been demonstrated as cross-calibratable to high precision (~ few percent) to similarly post-processed coronagraphic observations with STIS occulting wedges A and B.

8) Raw BAR5 coronagraphy has been demonstrated to reliably produce, with high repeatability, image contrasts of ~ $1 \times 10^{-3}$ @ 0.22", $3 \times 10^{-5}$ @ 1.0", $8 \times 10^{-6}$ @ 2.0", $2 \times 10^{-6}$ @ 3".

9) PSF-subtracted BAR5 coronagraphy has been demonstrated to 1-MAD contrast-limiting floors of ~ $4 \times 10^{-5}$ @ 0.22", $1 \times 10^{-6}$ @ 1.0", $1 \times 10^{-7}$ @ 1.5", $1 \times 10^{-8}$ @ $\geq 2$".

10) Raw BAR10 LL and LR coronagraphy has been demonstrated to produce raw IWA contrasts in the range of $10^{-3}$ to $4 \times 10^{-4}$ for targeted IWAs of 0.225" - 0.300", respectively, with statistically identical performance at r $\geq$ 0.4" – 0.5" regardless of IWA (see Fig. 18).

11) We recommend that BAR5 coronagraphy be resurrected from the grave and implemented as fully supported GO science capability.

## 16. Acknowledgements

We express our thanks to Charles Proffitt (STScI) in assisting us in understanding some of the technical details in our formulation of our Phase 2 observation and analysis plans.

This study is based on observations made with the NASA/ESA *Hubble* Space Telescope, obtained at the Space Telescope Science Institute (STScI), which is operated by the Association of Universities for Research in Astronomy (AURA), Inc., under NASA contract NAS 5-26555. These observations are associated with program #s 12923, 12551, 12228, and 11818. Support for program # 12923 was provided by NASA through a grant from STScI.

# Appendix A - Occulting BAR Metrology and Planned Image Scans

To initially construct the BAR10 corner, and cross-BAR5 image scans, we used the most contemporaneously available back-lit images of the coronagraphic image plane mask deployed by the STIS mode selector mechanism -- using instrumental flat fields, and images of Fomalhaut with its bright extended PSF halo back-lighting both occulters.

A crucial part of our observation planning was determining the precise POSTARG values to use for displacing the stars to the required occulter positions. *We were, however, unable to match the PDB SIAF.dat file values of the early-mission determined BAR10 position with high enough precision to a reference flat field image* (that we intended to use to determine the relative offsets needed). We thus decided to use archival BAR10 coronagraphic observations, combined with 50CORON flat field images, to find the BAR10 pixel coordinate for the specific flat field image we intended to use. The most contemporaneous suitable flat field we found was OBN409010, taken in 2011. We compared this to BAR10 occulted images of Fomalhaut taken in 2010 (P. Kalas, PI) from program GO 11818 (observations OB7914010 and OB7915010).

We determined the precise position of Fomalhaut in images from Visits 14 and 15 by finding the crossing point of the diffraction spikes at (OB7914010_raw_1.fits: 643.95 866.31, ob7915010_raw_1.fits: 644.07 866.83). We then determined the X and Y shift required to match the backlit silhouette of the BAR10 occulting element with its image on the flat field exposure (OB7914010_raw_1.fits: Flat Shift = +18.40, +19.30; OB7915010_raw_1.fits: Flat Shift = +18.40, +19.60). From these fits, we could infer the position of the nominal BAR10 position at 625.55 847.01 and 625.67 847.23. We took the average of these as the BAR10 pixel position on the flat field image:

Average (expected BAR10 pointing position) in FF image: 625.61 847.12 (in SIAF coordinates).

We noted that the position of Fomalhaut was offset by almost a full pixel in the image taken on orbit 13, which we discarded from our calculations. We also noted that the position of the BAR10 location appeared to be slightly asymmetrically placed w.r.t. the bar vertical axis of symmetry. As a result, positioning the star at the lower-left edge would place it 40.6 pixels from the initial BAR10 pointing, whereas positioning the star at the lower-right edge would be 41.8 pixels from the initial BAR10 pointing

We then determined the BAR10 LL and LR rounded corner locations. To do so, we measured the flat-field intensity profiles across the BAR10 rounded corners, at an angle of 45° w.r.t. the flat, orthogonal, bar edges. We thus ascertained the corner locations as the point at which the intensity was 50% of the background flat illumination level:

Lower Left:  SIAF(X, Y) = 598.125, 817.312
Lower Right: SIAF(X, Y) = 655.062, 817.380

For our first epoch (AU Mic) observations, we then defined image scans, across these positions offset by positions at 0.125", 0.15", 0.175", 0.2", 0.25", and 0.3" from these coroner locations along those 45° lines, as given in Table A-1.

From the same flat-field image we also determined the location of the center of the "unbent", but tilted, portion of BAR5 to be at SIAF(X,Y) = (970.06, 699.40). With that we also measured the line angle of long axis of the bar at 11.28° CW from image frame -X. We then defined BAR5 image scans including and about that central point, perpendicular to the along long axis offset by 0.00", ±0.04", and ±0.08", as also given in Table A-1.

Table A-1. AU Mic Image Scan Planned Locations*

```
wrt. edge   SIAF-X   SIAF-Y   POSTARGX   POSTARGY
LL_0.125:   599.868  819.055  -1.30538   -1.42318
LL_0.150:   600.217  819.404  -1.28770   -1.40550
LL_0.175:   600.565  819.752  -1.27002   -1.38782
LL_0.200:   600.914  820.101  -1.25234   -1.37014
LL_0.250:   601.611  820.798  -1.21699   -1.33479
LL_0.300:   602.308  821.495  -1.18163   -1.29943

LR_0.125:   653.319  819.123   1.40512   -1.41973
LR_0.150:   652.970  819.472   1.38744   -1.40205
LR_0.175:   652.622  819.820   1.36977   -1.38437
LR_0.200:   652.273  820.169   1.35209   -1.36669
LR_0.250:   651.576  820.866   1.31673   -1.33134
LR_0.300:   650.879  821.563   1.28138   -1.29598

B5+0.00:    970.060  699.400  17.46706   -7.49088
B5+0.04:    970.214  700.174  17.47488   -7.45165
B5+0.08:    970.369  700.947  17.48271   -7.41243
B5-0.04:    969.906  698.626  17.45924   -7.53011
B5-0.08:    969.751  697.853  17.45141   -7.56934
B5+0.12:    970.523  701.721  17.49053   -7.37320
B5-0.12:    969.597  697.079  17.44359   -7.60856
```
* POS TARGs with respect to the BAR10 aperture fiducial.

# Appendix B – Reduced Image Scan Positions and Inferred Pointing Corrections

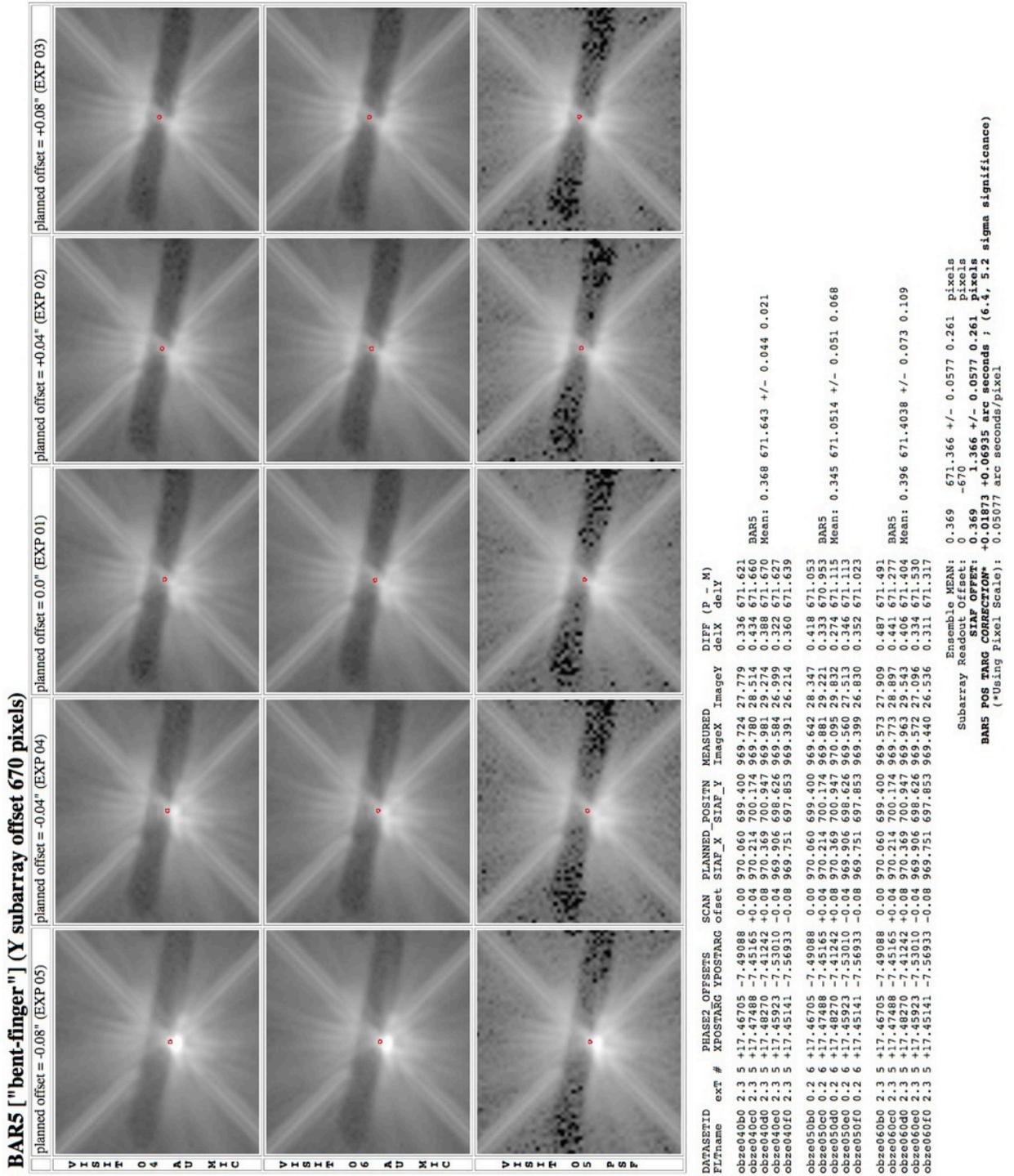

Figure B-1. AU Mic BAR5 image scans (Visits 04, 05, 06)
Table B-1. AU Mic BAR5 reduced positions, offsets, and errors

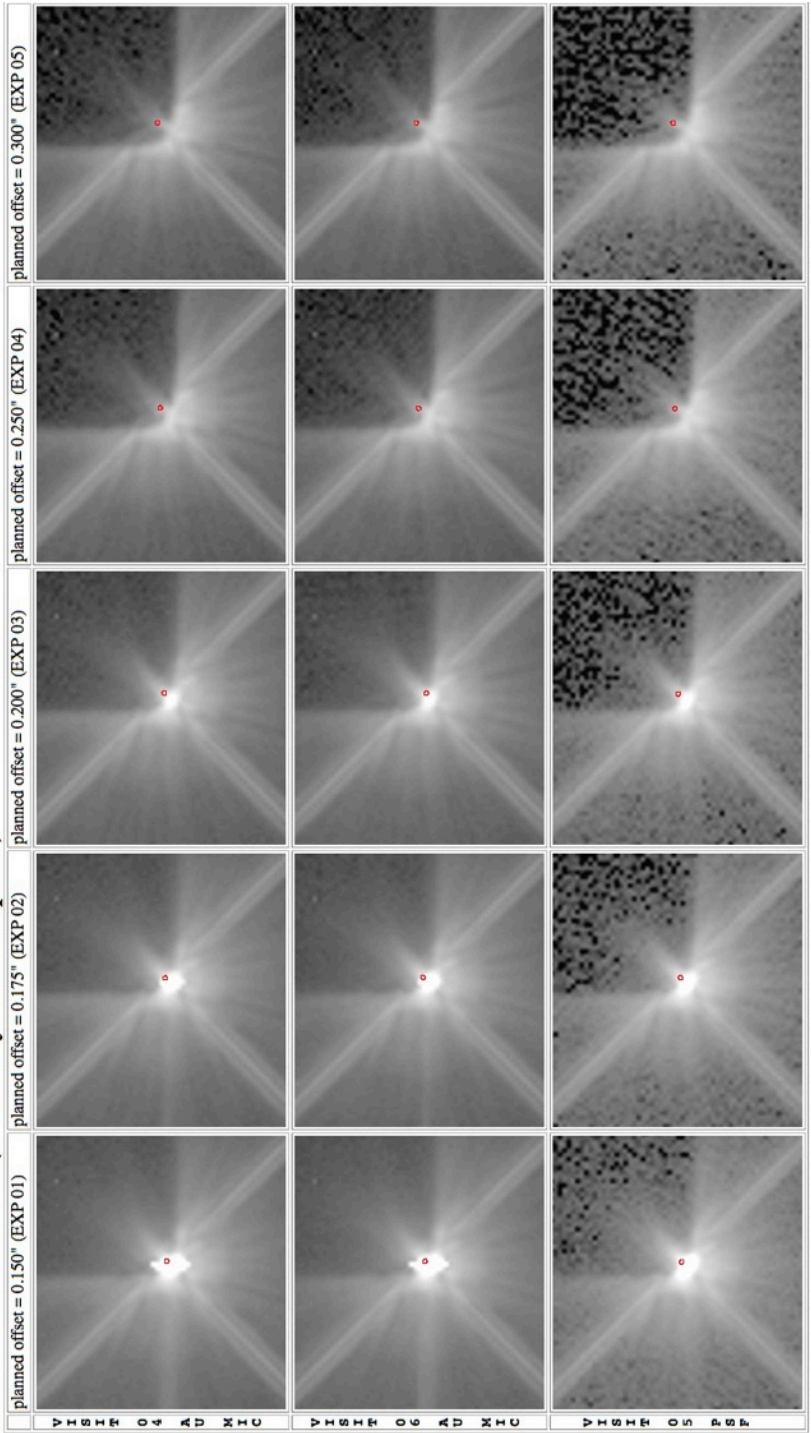

Figure B-2. AU Mic BAR10 LL image scans (Visits 04, 05, 06)
Table B-2. AU Mic BAR10 LL reduced positions, offsets, and errors

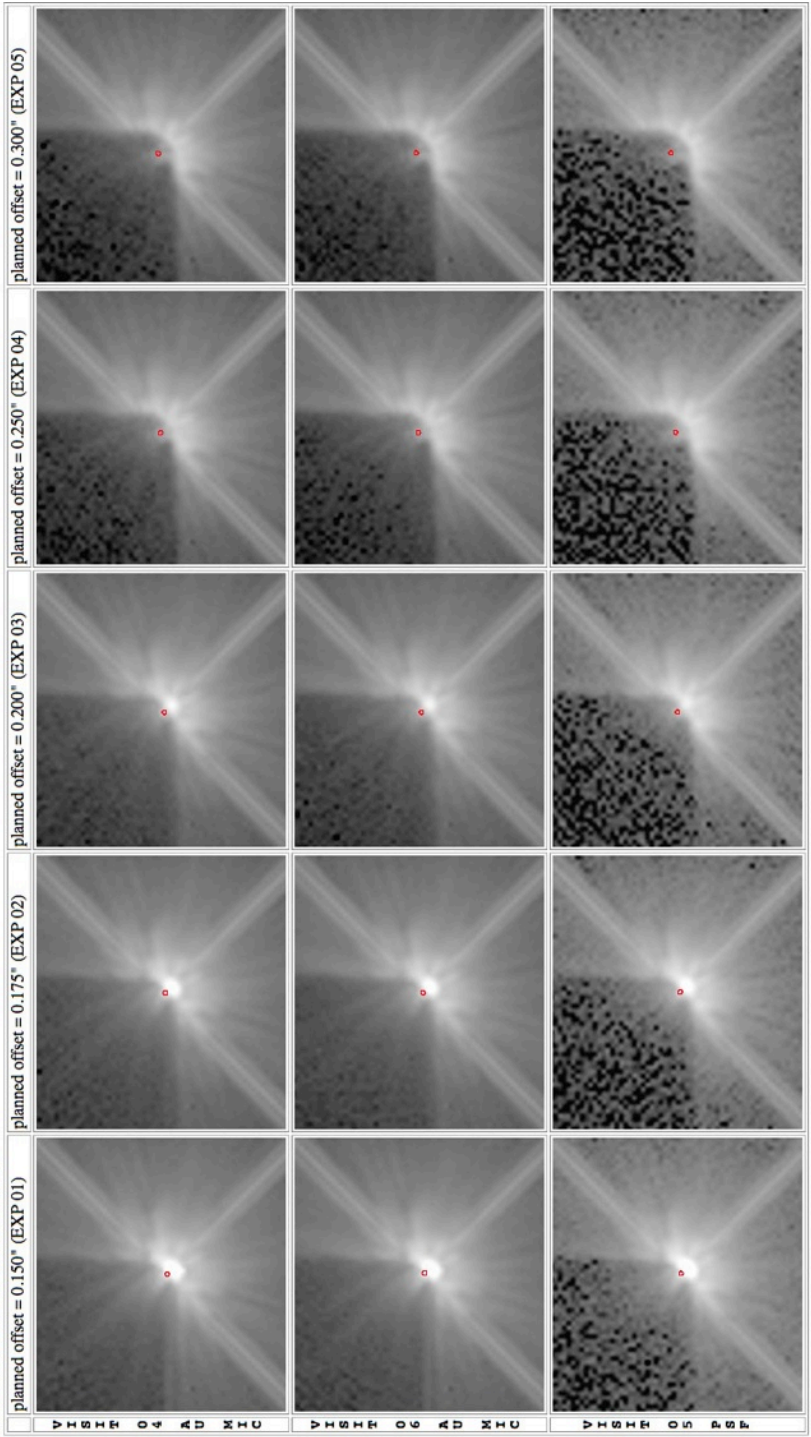

Figure B-3. AU Mic BAR10 LR image scans (Visits 04, 05, 06)
Table B-3. AU Mic BAR10 LR reduced positions, offsets, and errors